\documentclass[twocolumn,apj,appendixfloats,numberedappendix]{emulateapj-rtx4}

\usepackage{natbib}
\usepackage{float}
\usepackage{graphicx}
\usepackage{amsmath,amssymb}
\usepackage{ulem}
\usepackage{txfonts}
\usepackage{comment}

\newcommand{\vect}[1]{{\boldsymbol{#1}}}

\newcommand{\be}{\begin{equation}}
\newcommand{\ee}{\end{equation}}

\begin{document}
% Use the \preprint command to place your local institutional report
% number in the upper righthand corner of the title page in preprint mode.
% Multiple \preprint commands are allowed.
% Use the 'preprintnumbers' class option to override journal defaults
% to display numbers if necessary
%\preprint{}

%Title of paper
\title{Evolution of turbulence in the expanding solar wind, a numerical
study} 
%\title{Solar wind radial expansion: impact on the turbulent cascade}
\author{Yue~Dong}
\affil{LPP, Ecole Polytechnique, 91128 Palaiseau, France.}
\email{Yue.Dong@lpp.polytechnique.fr}

\author{Andrea~Verdini}
\affil{Dipartimento di Fisica e Astronomia, Universit\`a degli studi di
Firenze, Firenze, Italy;\\
SIDC-STCE, Royal Observatory of Belgium, Bruxelles, Belgium.}
\email{verdini@arcetri.astro.it}

\author{Roland~Grappin}
\affil{Ecole Polytechnique, 91128 Palaiseau, France.}
\email{grappin@lpp.polytechnique.fr}

\date{\today}

\begin{abstract}
We study the evolution of turbulence in the solar wind by solving numerically the full 3D magneto-hydrodynamic (MHD) equations embedded in a radial mean wind.
The corresponding equations (expanding box model or EBM) have been considered
earlier but never integrated in 3D simulations. Here, we follow the
development of turbulence 
from $0.2~AU$ up to about $1.5~AU$. Starting with isotropic
spectra scaling as $k^{-1}$, we observe a steepening toward a $k^{-5/3}$
scaling in the middle of the wavenumber range and formation of spectral
anisotropies.
The advection of a plasma volume by the expanding solar wind causes a non-trivial
stretching of the volume in directions transverse to radial and the
selective decay of the components of velocity and magnetic
fluctuations. 
These two effects combine to yield the following results.
(i) Spectral anisotropy: gyrotropy is broken, and the radial wavevectors have
most of the power.
(ii) Coherent structures: radial streams emerge that resemble the observed microjets.
(iii) Energy spectra per component: they show an ordering in good agreement with the one observed in the solar wind at 1 AU.
The latter point includes a global dominance of the magnetic energy over kinetic energy in the inertial and $f^{-1}$ range and a dominance of the perpendicular-to-the-radial components over the radial components in the inertial range. 
We conclude that many of the above properties are the result of evolution during transport in the heliosphere, and not just the remnant of the initial turbulence close to the Sun.
%Direct comparison with solar wind data at 1 AU confirms that these 
% two basic phenomena are at work in solar wind turbulence.
%We also measure the distribution of the decay rate with distance, an important step towards understanding the large injection scales in the solar wind in the various directions.
\end{abstract}

% insert suggested PACS numbers in braces on next line
%\pacs{47.65.+a, 47.27.Eq, 47.27.Gs}
% insert suggested keywords - APS authors don't need to do this
%\keywords{}
%\maketitle must follow title, authors, abstract, \pacs, and \keywords
\maketitle

%*************************************************************
\section{Introduction}

Since the early observations by \citet{1968ApJ...153..371C} and \citet{Belcher:1971cn}, 
solar wind turbulence has been considered a good example of well-developed MHD turbulence, with power-law spectra occupying a large frequency range.
This turbulence shows specific features: (i) an important contribution of coherent structures in the spectrum (\citet{Bruno:2007tw}; \citet{1993JGR....98.1257T}), 
(ii) a large cross-helicity in the fast streams far from the heliospheric
current sheet (iii), a non-trivial ordering of the different components \citep{Belcher:1971cn}, and (iv) a large magnetic dominance in slow streams \citep{Grappin:1991tr}. 

Turbulent dissipation is one of the most important consequences of a turbulent cascade and a test for any theory. In homogeneous, stationary turbulence, it is equal to the energy injected in the system and the energy taken out of it, that is in turn equal to the heating rate.
Turbulent dissipation should be equal as well to the energy flux transmitted
from one scale to the other along the inertial range, that is, the scale range
of the turbulent cascade, which is given by the third moment of the fluctuations. The latter quantity has indeed been measured and sometimes found to be scale-independent \citep{2008ApJ...679.1644M,2007PhRvL..99k5001S}, but an agreement with different possible evaluations of the injection rate (that depends on the turbulence theory) is still a matter of debate \citep{2007JGRA..11207101V}.

A most important tool to analyze a turbulent spectrum is to look at power-law scaling in the second order moment of fluctuation.
Most observational data are made of 1D spectra that are obtained collecting
records along the radial direction. An important exception is provided by the
Cluster spacecrafts that allow to recover 3D information via the use of k-filtering theory
\citealt{2010PhRvL.105m1101S}, \citealt{Narita_al_2010}), 
but this is limited to relatively large scales and at a distance of 1~AU.
Knowledge on the 3D structure of the solar wind fluctuations is however essential for several reasons:
(i) theories based on resonant interactions are 
strongly anisotropic, the spectrum being more and more confined into the plane
perpendicular to the local mean magnetic field, as the cascade proceeds; 
(ii) the 3D structure has
implications on the energy injection rate and hence on the associated
dissipation.

Single spacecraft measurements can still be used to reconstruct the 3D structure by adopting a strong
hypothesis, i.e. gyrotropy around the mean magnetic field axis. Exploiting the wandering of the magnetic field direction, one can recover the structures of turbulence in the field-parallel and field-perpendicular directions \citep{Matthaeus_al_1990,Dasso_al_2005}.

In the solar wind, as we will see, expansion itself is a basic source of component anisotropy and spectral anisotropy that are \textit{fed into the cascade range}. Here, we recall the basics \citep{GVM93}, 
(i) the component anisotropy is forced by expansion into
the turbulent system as a consequence of the conservation of linear invariants
like, mass, radial momentum, angular momentum, and magnetic flux, (ii) The
spectral anisotropy is also forced into the system due to the kinematic stretching of the whole plasma volume in the two directions transverse to the mean radial wind.

%There are two basic ways to describe the evolution of the turbulent plasma, the
%eulerian and the lagrangian descriptions.
A first category of models that account, to some extent, for the effects of
expansion is provided by transport models with strongly simplified nonlinear
interactions
\citep{Tu_al_1984,1987SoPh..109..149T,Velli:1993to,1990CoPhC..59..153V,Matthaeus:1999p630,CB05,2007ApJS..171..520C,Verdini_Velli_2007,Verdini:2010et,Chandran:2011p2723,Lionello:2014p3023}.
A second category describes the turbulent evolution using a 
detailed model of turbulence, Reduced MHD or shell-reduced MHD  
\citep{Verdini:2009ih,2012ApJ...750L..33V,Perez:2013p3017}. 
%In these works the back-reaction of turbulence on the wind is neglected, and the integration domain is close to the Sun, in the acceleration region.
However, even the most complete of the previous descriptions, that is the
Reduced MHD model, lacks essential ingredients. 
Reduced-MHD equations are obtained in the limit of a strong mean field, which
is a reasonable assumption in the accelerating region of the solar wind.
This leads to quasi-2D spectra with no parallel fluctuations.
At larger heliocentric distances, this approximation is too restrictive
as we will show, and there are indications that this is also true in the
acceleration region \citep{Matsumoto:2012p2665}.
In addition, a large amount of the computer memory is devoted to the
description of the stratification, thus limiting the resolution and hence
the Reynolds number that are achievable. This issue becomes even more important
when dealing with system size of the order of one astronomical unit, as we do.

In the present work we will adopt a pseudo-Lagrangian description, which
allows us to get rid of restrictions on spectral and component anisotropy
and to take full advantage of available computer resources in describing
turbulence.
This is made possible in the Expanding Box Model (EBM) since all the above
effects of expansion are incorporated in the original MHD equations as linear, time-dependent terms.

%The theoretical frame work that allows to include the effects of the expansion while following a plasma volume embedded in a given radial wind
%without spoiling the nonlinear machinery is called the expanding box model (EBM). 
 
The EBM was introduced and used previously in the 1D and 2D cases to account for the anisotropic nature of expansion, in the works by \citet{GVM93, Grappin_Velli_1996,Grappin_1996}.
These early results may be summarized as follows. First, the transverse wind expansion disrupts a nonlinear Alfv\'en wave with constant magnetic pressure,
due to the progressive rotation in opposite directions of the wave vector and mean guide field.
Second, the combined effects of the expansion and shear due to the stream structure lead to a decrease of 
Alfv\'enicity comparable to that observed in the Solar Wind.
Finally, the formation of small scales is largely inhibited by the expansion: 
the solar wind turbulent spectrum should thus have formed early in the corona, not in the solar wind itself.
The scope of these early studies was however limited as it dealt with 2D MHD.
Other works dealing with the parametric instability using the EBM (in the MHD or hybrid version) are to be found in \cite{Tenerani:2013p3022,Matteini:2006jn}.

In the present work, we use the 3D EBM to explain a number of features related to anisotropy both in simulations and the solar wind: (i) the component anisotropy (ii) the spectral anisotropy.

The plan is as follows. In Section 2, we describe the expanding box
model and enumerate the basic conservation laws verified by the
model. The results are reported in Section 3. We first consider
expansion with no mean field; it allows to introduce the new effects
of expansion without the complications of a second symmetry axis. We
next deal in detail with a mean field aligned with the radial (that
remains aligned during the radial evolution). This case combines all
physical effects, albeit in a relatively simple geometry, since the
two symmetry axes are aligned. We finally consider the realistic case
with an oblique magnetic field, in which the magnetic field follows
the Parker spiral. 
These results are compared with observations in the
Discussion. 
%The effect of a rotating mean field has been
%studied by Ghosh (), but never as an autocoherent consequence on the
%expansion. The last section is a discussion.

%---------------------------------------------------------
\begin{figure}%[t]
\begin{center}
\includegraphics [width=0.49\linewidth]{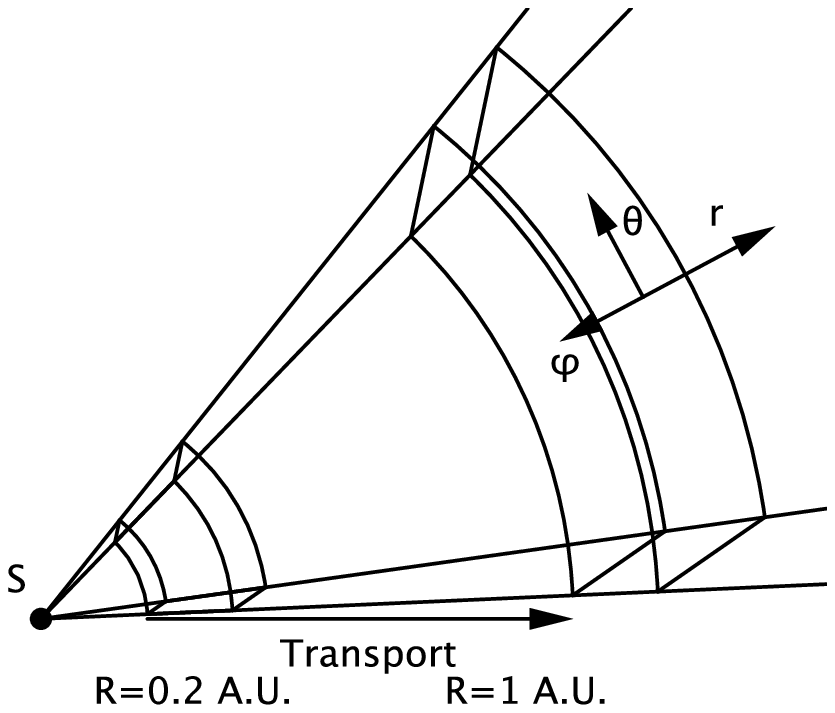}
\includegraphics [width=0.49\linewidth]{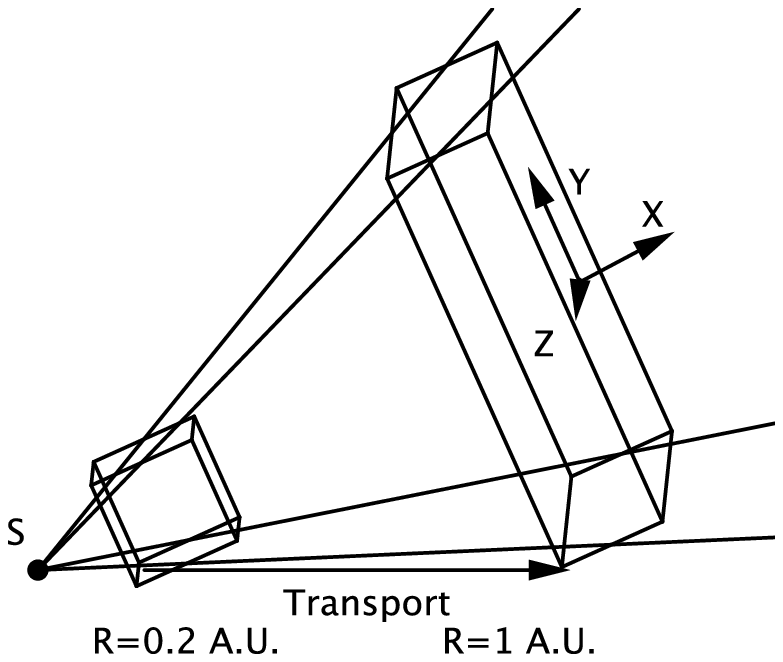}
\caption{
A volume of plasma transported by a radial wind with uniform speed. 
The volume expands in the transverse directions but not along the
radial direction. Left: spherical expansion. Right: comobile approximation.
}
\label{fig1}
\end{center} 
\end{figure}
%---------------------------------------------------------

\section{Equations, parameters and basic physics}
\label{sec:expanding-box-model}

\subsection{Expanding Box Model}
We give here a short derivation of the Expanding Box model (see
\citealt{GVM93,Grappin_Velli_1996,Rappazzo_al_2005}.
The wind is assumed to be radial and to have uniform velocity ($U_0=const$). 
The radius $R$ at which the box is located varies with time $T$ as:
\be
R(T) = R_0 + U_0 T
\ee
where $R_0$ is the initial position of the box.
We write now the equations in adimensional form.
Space, time and velocity are measured in the following units:
\begin{align}
\mathcal{L}= L_0/(2\pi)\\
\mathcal{T} = t_{NL}^0=L_0/(2\pi u_{rms}^0) \\
\mathcal{U} = u_{rms}^0
\end{align}
where $u_{rms}^0$ is the initial rms velocity of the fluctuations, and $t_{NL}^0$ is the initial nonlinear time based on the initial $rms$ velocity, and $L_0$ is the thickness of the box.

A first possibility to follow the evolution of the plasma is to use a spherical coordinate system centered on the Sun ((cf. Fig.~\ref{fig1}a). We prefer to use the simpler Cartesian coordinates.
Consider a Cartesian frame with $x$ axis parallel to the radial passing through the middle of the box,
change to the Galilean frame moving with the mean wind along the radial
coordinate, $X'=X-U_0 T = X - (R(T)-R_0)$. 
In this frame, an initially cubic box is uniformly stretched in the transverse directions and becomes a parallelepiped of aspect ratio (cf. Fig.~\ref{fig1}a and \ref{fig1}b):
\be
a(T)= R(T)/R_0.
\ee
If we write down the MHD equations for the fluctuating velocity at this stage, a
new term appears. Such term is proportional to $U_\bot \nabla$ and describes the systematic advection of the plasma in the directions transverse to the radial by the mean flow after the Galilean frame change.
This term disappears from the equations if we move now to \textit{comobile} coordinates $t,x,y,z$:
\begin{align}
t = T \nonumber \\
 x = X' \nonumber = X - U_0 T\\
 y = Y/a(t) \nonumber \\
 z = Z/a(t) 
\label{cartesian}
\end{align}
Suppressing the advection by the transverse flow allows to assume periodicity
of all fields expressed as functions of the comobile coordinates: density
$\rho$, pressure $P$, magnetic field $\mathbf B$, and velocity fluctuation
$\mathbf U = \mathbf V - U_0 \mathbf{\hat e_r}$, where $\mathbf V$ is the total velocity of the wind.
Periodicity is a basic requirement, as otherwise it would be impossible to
determine the boundary conditions, as we have no \textit{a priori} information of the plasma outside the volume considered.

The comobile coordinate system is in fact equivalent to spherical coordinates centered on the Sun, but with the radial coordinate measured locally with respect to the local Galilean frame.
This is valid if the thickness of the box $L_0$ is always small compared to the heliocentric distance $R$:
\be
L_0/R(t) \ll 1.
\label{valid}
\ee
This assumption allows us to (i) neglect curvature terms in the previous derivation and (ii) to assume periodicity in the radial direction as well.
Omitting dissipation terms, the equations take the form of standard MHD
equations, with however two modifications, (i) additional linear terms involving
the mean velocity $U_{0\bot}$  appear in the right-hand side (ii) a new expression
for the gradients is used, accounting for the lateral stretching:
\begin{flalign}
&D_t \rho +\rho \nabla\cdot \vect{U} = - \rho \nabla \cdot \vect{U_{0\bot}} =
 -2  \rho \frac{\epsilon}{a}
\label{eqrho} \\
&D_t \vect{U} + \frac{1}{\rho} (\nabla
(P+\frac{B^2}{2}) - \vect{B}\cdot\nabla \vect{B}) = - \vect{U} \cdot \nabla \vect{U_{0\bot}}=
-\alpha \vect{U} \frac{\epsilon}{a} 
\label{equ} \\
&D_t \vect{B} 
- \vect{B}\cdot\nabla \vect{U} +
\vect{B} \nabla\cdot\vect{U} = - \vect{B} \nabla \cdot \vect{U_{0\bot}} + \vect{B} \cdot \nabla \vect{U_{0\bot}} = 
 -\beta \vect{B} \frac{\epsilon}{a}
\label{eqB} \\
&D_t P + \frac{5}{3} P \nabla\cdot\vect{U} = - \frac{5}{3} P \nabla \cdot \vect{U_{0\bot}} = 
-\frac{10}{3} P \frac{\epsilon}{a} 
\label{eqP}\\
&P = \rho \ T_p \\
&\nabla = (\partial_x,(1/a)\partial_y,(1/a)\partial_z) 
\label{gradperp} \\
&a(t) = 1 + \epsilon t
\label{aratio}
\end{flalign}

The two coefficients $\alpha = (0,1,1)$ and $\beta = (2,1,1)$ produce a
differential damping of the components $x, y, z$ of the magnetic and kinetic
fluctuations. 
Along with usual MHD non-dimensional parameters, like the sonic and Alfv\'enic Mach
number, the EBM equations are ruled by expansion parameter $\epsilon$ that appears in the rhs of the evolution equations:
\begin{equation}
\epsilon=\frac{U_0 L_0}{2\pi u^0_{rms} R_0}=\frac{t_{NL}^0}{t_{exp}^0}
\label{epsi}
\end{equation}
%%where $L_0$ and $R_0$ are respectively 
%the initial size and position of the box, and $u^0_{rms}$ is the rms value of the
%fluctuations at the initial time 
In the following we will take $L_0=R_0$. 
Thus, $t_{exp}^0 = R_0/U_0$ is the initial expansion time and $t_{NL}^0$ is the initial nonlinear time.
Note that the expansion time may also be called transport time, as it is the time necessary to transport the plasma from the Sun to the distance $R_0$ at the constant velocity $U_0$.
Because of this, the inverse of the expansion parameter is also called the ``age'' of the turbulence, being the transport time expressed in units of nonlinear times \citep{Grappin:1991tr}:
\be
Age = t_{exp}^0/t_{NL}^0 = 1/\epsilon
\label{age}
\ee

Note that, while the equations are written in terms of comobile coordinates, the numerical results will be presented in the standard physical cartesian coordinate system, in real space as well as in Fourier space.

We have omitted the dissipative terms in the equations of the model. We give them below (omitting all other terms) in each evolution equation:
\begin{eqnarray}
\partial_t U &=& \nu (\tilde \nabla^2 U + \frac{1}{3} \tilde \nabla (\tilde\nabla \cdot U))
\label{nudec0}
\\
\partial_t B &=& \eta \tilde\nabla^2 B
\\
\tilde \nabla &=& (\partial_x,\partial_y,\partial_z)
\\
\nu &=& \eta =  \nu_0 /a(t)
\label{nudec}
\end{eqnarray}
where $a(t)= R(t)/R_0$ is the normalized heliocentric distance (eq.~\ref{aratio}).
These expressions differ from the usual ones in several ways. 
The viscosity $\nu$ is kinematic (independent of the density $\rho$), the derivation operators are defined with respect to comobile coordinates and
the transport coefficients $\nu$ and $\eta$ decrease with heliocentric distance. 
These choices are all dictated by the same goal: maintaining a substantial Reynolds number during the integration.

First, we considered a kinematic viscosity since with a dynamic viscosity a
factor $1/\rho\propto R^2$ would
result in a drastic damping of the energy, which we want to avoid.
Second, the gradient operators appearing in these equations should be the 
physical nablas defined in eq.~\ref{gradperp}. 
However, in view of the limited Reynolds number achievable in direct
numerical simulation, the increase of all physical characteristic scales in directions perpendicular to the radial would lead to a too strong damping of all fluctuations perpendicular to radial. 
Finally, the linear damping of energy with distance due to expansion that
sums up to the usual turbulent damping would lead to a drastic drop of the
Reynolds number. The systematic decrease of the transport coefficients with
distance in eq.~\ref{nudec} is a prescription that compensates for the above
effects and maintains the Reynolds number at a reasonable level.
The price to pay for this modification of the standard viscous terms is that the heating cannot be calculated 
in a self-consistent way.

Indeed, the dissipative terms just described must in principle have their counterpart appearing in the energy equation. The standard conservative expression for the homogeneous MHD is (omitting the adiabatic terms)
\be
\partial_t P = \frac{2}{3} (\mu (\omega^2 +1/3 (\nabla \cdot u)^2)+ \eta J^2 + \kappa \nabla^2 T_p)
\label{heat}
\ee
In view of the modifications of the dissipative terms described in eqs.~\ref{nudec0}-\ref{nudec},
no simple modification of the heating terms in eq.~\ref{heat} can be found, that would correctly express the
transfer to the internal energy (i.e., heating) of the energy flux lost by the fluctuations at small scales by the dissipative terms.
We thus have chosen here to minimize the heating due to expansion: in the present model, the rhs terms in eq.~\ref{heat} have been divided by the density for the viscous term, with the resistive and conductive terms being divided by the average density. 
As a consequence, in all runs with expansion, the resulting irreversible heating has proved to be negligible compared to the $R^{-4/3}$ cooling resulting from the plain adiabatic expansion (eq.~\ref{eqP}).

We postpone for future work finding expressions of the dissipative terms and the heating terms that are truly conservative and at the same time prevent the sharp drop of the Reynolds number that would inevitably result from adopting standard dissipative terms.

\subsection{Energy spectra and characteristic times}

In the following we use $k$ to denote wavevectors in comobile
Fourier space, and the capital $K$ to denote wavevectors in the
physical Fourier space ($K_x=k_x,~K_{y,z}=k_{y,z}/a$). 
Consider now a scalar quantity $\psi$. The Fourier coefficient $\hat\psi$ is defined as:
\be
  \label{eq:fourier_def}
  \hat \psi(\vect{K}) = \frac{1}{V}\int e^{-i \vect{K} \cdot \vect{X} } \psi(\vect{X}) d^3\vect{X} \\
\ee
and the 3D spectral density:
\be
  E_{3D}(\vect{K}) = \mid \hat \psi(\vect{K}) \mid ^2
\ee
so that total energy verifies:
\begin{equation}
  \label{eq:como_fourier_en}
  2 E_\psi= \int E_{3D}d^3\vect{K}
  = \frac{1}{V}\int |\psi(\vect{x})|^2 d^3x = \psi_{rms}^2
\end{equation}
%%%%%%%%

%%%%

We further define the reduced 1D spectra along the radial direction $E_{1D}(k_x)$ 
and the gyrotropic spectra $E_{3D}^{gyro}$ (analogous definition hold in
comobile space once $K$ is replaced by $k$):
\begin{eqnarray}
E_{1D}(K_x)&=&\int E_{3D}(K_x,K_y,K_z) dK_ydK_z 
\label{e1d}\\
E_{3D}^{gyro}(K_{x},K_\bot)&=&\frac{1}{2\pi}\int E_{3D}(K_{x},K_\bot,\phi)d\phi
\end{eqnarray}
We use the reduced 1D spectra (summing spectra on the three components) to define the velocity field $U(K)$ 
\begin{eqnarray}
%u(k_x)=\sqrt{k_xE_{1D}(k_x)}\\
%u(k_{||},k_\bot)=\sqrt{(k_{||}^2+k_\bot^2)^{3/2}E_{3D}^{gyro}(k_{||},k_\bot)}
U(K_x)&=&\sqrt{K_xE_{1D}(K_x)}
%\\
%u(\mathbf{K})&=&\sqrt{K^{3}E_{3D}(\mathbf{K})}
\end{eqnarray}
that enters in the definition of the nonlinear time at a given wavenumber,
\be
t_{NL}=(KU(K))^{-1}
\label{eq:characteristictimes_tnl}
\ee
The Alfv\'en and expansion times are further defined as
\begin{eqnarray}
\label{eq:characteristictimes_te}
t_{exp}&=&R(t)/U_0 \\
t_A&=&((\vect{B_0}/\sqrt{\bar{\rho}})\cdot\vect{K})^{-1}
\label{eq:characteristictimes_ta}
%t_\nu&=&1/{K}^2\nu\\
\end{eqnarray}

%*************************************************************
%------------------------------------------------------------
\subsection{Basic physics}
We describe below the three kinds of damping that affect the turbulence evolution in a spherically expanding flow.
\subsubsection{Linear expansion (no Alfv\'en coupling)}\label{LEE}
The right hand sides of eqs.~\eqref{eqrho}-\eqref{eqP} imply the
conservation of mass, angular momentum, and magnetic flux, ($\bar{\rho }\propto
1/R^2$, $U_y,U_z\propto1/R$, $B_x\propto1/R^2$, $B_y,B_z\propto1/R$) while
eq.~\eqref{eqP}
accounts for the adiabatic temperature decrease in absence of heating ($T_p \propto
R^{-4/3}$).
The conservation laws for velocity and magnetic field are best expressed in terms of velocity units \citep{Grappin_Velli_1996}:
\begin{eqnarray}
B_x/\sqrt{\bar{\rho}} &\propto& 1/R \label{exp1}\\
B_{y,z}/\sqrt{\bar{\rho}} &\propto& 1 \label{exp2}\\
U_x &\propto& 1 \label{exp3}\\
U_{y,z} &\propto& 1/R
\label{exp4}
\end{eqnarray}
Such decay laws hold for average properties.
They also apply to the amplitudes of the fluctuations at the largest scales,
where nonlinear interactions and the coupling with the mean field are negligible, in other words when:
\be
t_{exp} \le (t_{NL}, t_A)
\ee
that define scales belonging to a regime that we will term as non-WKB regime.
\subsubsection{Linear expansion with Alfv\'en coupling}
When a mean magnetic field is present and strong enough, the kinetic and magnetic fluctuations become coupled and one expects the previous damping laws to be modified \citep{Grappin_Velli_1996}.
The expected decay law in that case is what is usually called the WKB law, with
a scaling intermediate between those of the different components described in eqs.~\ref{exp1}-\ref{exp4}:
\be
B_{i}(k_{||B0})/\sqrt{\bar{\rho}} = U_{i}(k_{||B0})\propto 1/R(t)^{1/2}\;\;\;\mbox{for}~i=x,y,z
\label{wkb}
\ee
The condition for strong Alfv\'enic coupling (WKB) is
\be
t_A<t_{exp}
\ee

We will call ``non-WKB'' the regime where the different components are decoupled and ``WKB'' the regime with Alfv\'en coupling.
The WKB decay is valid in principle at scales small enough, for wave vectors along the meal field, while the non-WKB decay holds at larger scales.
We define $K_A$ as the critical wavenumber that divides the spectrum into two branches obeying respectively the two decaying laws,
\be
K_A = U_0/(|\vect{B_0}/\sqrt{\bar{\rho}}|R)
\ee
As the mean magnetic field $B_0$ rotates with time/distance due to
the conservation of magnetic flux ($B_{0x,y,z}/\sqrt {\bar{\rho}} \propto
(1/R,1,1)$), the critical wavenumber $K_A$ will remain constant in the radial
direction but will decrease as $1/R$ along the perpendicular $K_{y,z}$
directions. 

Alfv\'en coupling will play an active role even when $B_0=0$.
This Alfv\'en coupling is weaker than the one built on the global mean field, as it is based on the smaller \textit{local} mean field (typically of order $b_{rms}$), yet it will have important effects.

\subsubsection{Nonlinear damping}

In addition to the linear damping caused by expansion, one must consider the turbulent damping that occurs at intermediate-small
scales where the nonlinear time is small enough.
Turbulent dissipation is due to the viscous and resistive terms operating at very small
scales that are feeded by the cascade process.
The wavevector 
\be
K_{exp}=U_0/(RU(\mathbf{K}))
\ee
for which $t_{exp}=t_{nl}$ divides in principle
scales with a decay dominated by expansion ($K<K_{exp}$) and cascade ($K>K_{exp}$).
It defines a surface in 3D Fourier space that can expand or contract in time, depending on the direction (radial or transverse), and depending on whether the effective decay of $u(\mathbf{K})$ 
(i.e., including the linear and the nonlinear decay) is faster or not than $1/R$. 

Such decay cannot be easily predicted since wavevectors are coupled by the
cascade, while non-WKB and WKB decay are at work at different large scales and
at different wavevector orientations.
%Note that expansion causes a reduction of nonlinear interactions
%when wavevectors perpendicular to the radial are involved, since the stretching
%of the box implies $k_y,k_z\propto1/R$. 

%------------------------------------------------------------
\subsection{Numerics - Initial conditions and parameters \label{sec-param}}
The spatial scheme is pseudo-spectral, that is, the gradients are computed in Fourier space by plain multiplication by wavevector components. At each time step, an isotropic truncation is done in comobile coordinates: all modes with $k>k_{max}=N/2$ are eliminated, with N the number of grid points in each direction. The temporal scheme is Runge-Kutta of order 3.

The initial conditions are as follows. 
Rms velocity and magnetic fluctuations are unity; density and temperature are uniform. 
The corresponding Mach number is substantially smaller than unity (see later eq.~\ref{mach}):
\begin{align}
\rho = 1 \\
T_p = 40 \\
u_{rms}\sim b_{rms}\sim 1
\end{align}
The initial fluctuations are a superposition of incompressible fluctuations
with random phases that form an isotropic spectrum 
at equipartition between kinetic and magnetic energy.
There is no correlation between the velocity and magnetic field fluctuations.
The spectrum for $u$ or $B$ has the form $E_{3D}(\vect{k}) \propto
| \vect{k} | ^{-3}$ 
that reduces to a 1D spectrum $\propto k^{-1}$.

There are two free parameters, the initial expansion rate $\epsilon$, and the mean magnetic field $B_0$.

$\epsilon$ is the
ratio of non-linear over expansion times computed at the largest scale $L_0$
at initial distance $R_0$ (eq.~\ref{epsi}). 
In order to have a chance to observe significant effects of the expansion on
the spectral evolution, we choose the value $\epsilon = 2$ in our ``expanding'' runs.
\begin{table}
\centering
\caption{List of runs and parameters.
$B_0=(B_{0x},B_{0y})$ is the mean magnetic field in Alfv\'en units at $t=0$, $\epsilon$ is the
expansion parameter, $N$ is the grid point number, $t$ and $R/R_0$ are the simulation duration and final aspect ratio of the domain, $\nu_0$ is the initial viscosity (see eq.~\ref{nudec}).
}
\begin{tabular}{ccccccccc}
\hline
run & $B_0$ & $\epsilon$ & $N$ & t & $R/R_0$ & $\nu_0$ & name\\
\hline\hline
A   & (0, 0)      & 0 & 512  & 4   & 1 & $2 10^{-4}$& FY28B0E0   \\
B   & (0, 0)      & 2 & 512  & 3.2  & 7.4 & $10^{-4}$& FY34B0E2   \\
C   & (2, 0)      & 0 & 512  & 4 & 1 &$10^{-4}$&FY34BR2E0  \\
D   & (2, 0)      & 2 & 512  & 4   &9 &$2 10^{-4}$& FY28BR2E2  \\
E   & (2, 2/5)    & 2 & 1024 & 1.8 &4.6 &$2 10^{-4}$ & FY13B2E2   \\
%S & Slow         & (1, 1/5)    & 1 & 512  & 4   & FY33B1E1   \\
%W & Weak         & (1/2, 1/10) & 2 & 512  & 4   & FY30B0P5E2 \\
\hline\hline
\end{tabular}
\label{table1}
\end{table}

In order to identify the plasma scale that is submitted to such an expansion
parameter, we examine the values of $\epsilon$ (actually its inverse, the
``age'' of the plasma) that have been measured in Helios data at different
heliocentric distances \citep{Grappin:1991tr}. On average $\epsilon=2$ for the day-scale, and $\epsilon=0.1$ for the hour-scale.
The time scale is related to the spatial scale $L$ by the Doppler relation $\tau = L/U_0$, $U_0$ being the bulk wind speed. 
Assuming $U_0=600$ km/s, one thus finds that a volume of radial size $L=5 \ 10^7$~km $\simeq 0.3 AU$ is subject to an expansion rate $\epsilon \simeq 2$.

Our starting heliocentric distance will be $R_0=0.2 AU$, thus the initial
plasma volume almost touches the Sun. Hence our computational box 
has a large angular size, implying that the curvature terms (neglected in the EBM equations) are important (cf. eq.~\ref{valid}).
Contrary to expectations, this has no deep consequences on the dynamical
evolution of the system as shown in \citet{Grappin_Velli_1996} who compared the
EBM with an eulerian simulation in an even more extreme (2D) configurations.

The second parameter is the mean magnetic field $B_0$. We adopt two values:
either $B_0=0$ or $B_0\approx2$. Since initially $\rho=1$, this is in Alfv\'en speed unit.
Two meaningful normalized numbers are the initial Mach number and Alfv\'enic Mach number; they are:
\begin{align}
M = u_{rms}/c_s = 0.12 \\
M_A = b_{rms}/B_0 = 0.5
\label{mach}
\end{align}

The low Mach number ensures that the compressible part of the velocity field will remain small.
Note that these parameters are not meant to be representative of fast winds.
Such a study needs to consider as well the effect of velocity-magnetic field correlation (which is high in the fast winds): this is postponed for a later study.

The runs of decaying turbulence that we analyze are listed in Table 1. 
Runs A and B are standard homogenous non-expanding runs ($\epsilon=0$),
respectively without or with a mean magnetic field (${\bf B_0}=B_0{\bf e_x}$).
Runs C and D are the corresponding expanding simulations ($\epsilon=2$).
The mean field $B_0$ is aligned to the radial direction in run D, while run E
has an initial oblique mean field that turns according to the Parker spiral
during the radial evolution.
Runs with mean field (C, D, and E) have $K_A<K_{exp}$ at the initial time, 
so for wavevectors along $B_0$ we will encounter first the
non-WKB decay at small wavenumbers, then the WKB decay at intermediate wavenumbers, and finally the turbulent-cascade decay at high wavenumbers.
%

%*************************************************************
\section{Results}
In the following we will consider first the overall properties of run B with expansion and no mean field, then consider for the four runs A, B, C, D the evolution with time/distance of total (kinetic + magnetic) energy and energy spectra. We will examine run E with an oblique magnetic field in the discussion section.

From now on, the magnetic field will be given in Alfv\'en
speed units, i.e., we will write $B$ for $B/\sqrt{\bar{\rho}}$
and time will be given in initial non-linear time unit $t_{NL}^0$.
Also, recall that the $x$ component of a vector field is a synonym for the radial component, while $y$ and $z$ components are components transverse to the radial.

\subsection{Emergence of structures in real space (run B)}
\label{sec:emerg-struct-real}
\begin{figure}[t]
\begin{center}
\includegraphics [width=0.97\linewidth]{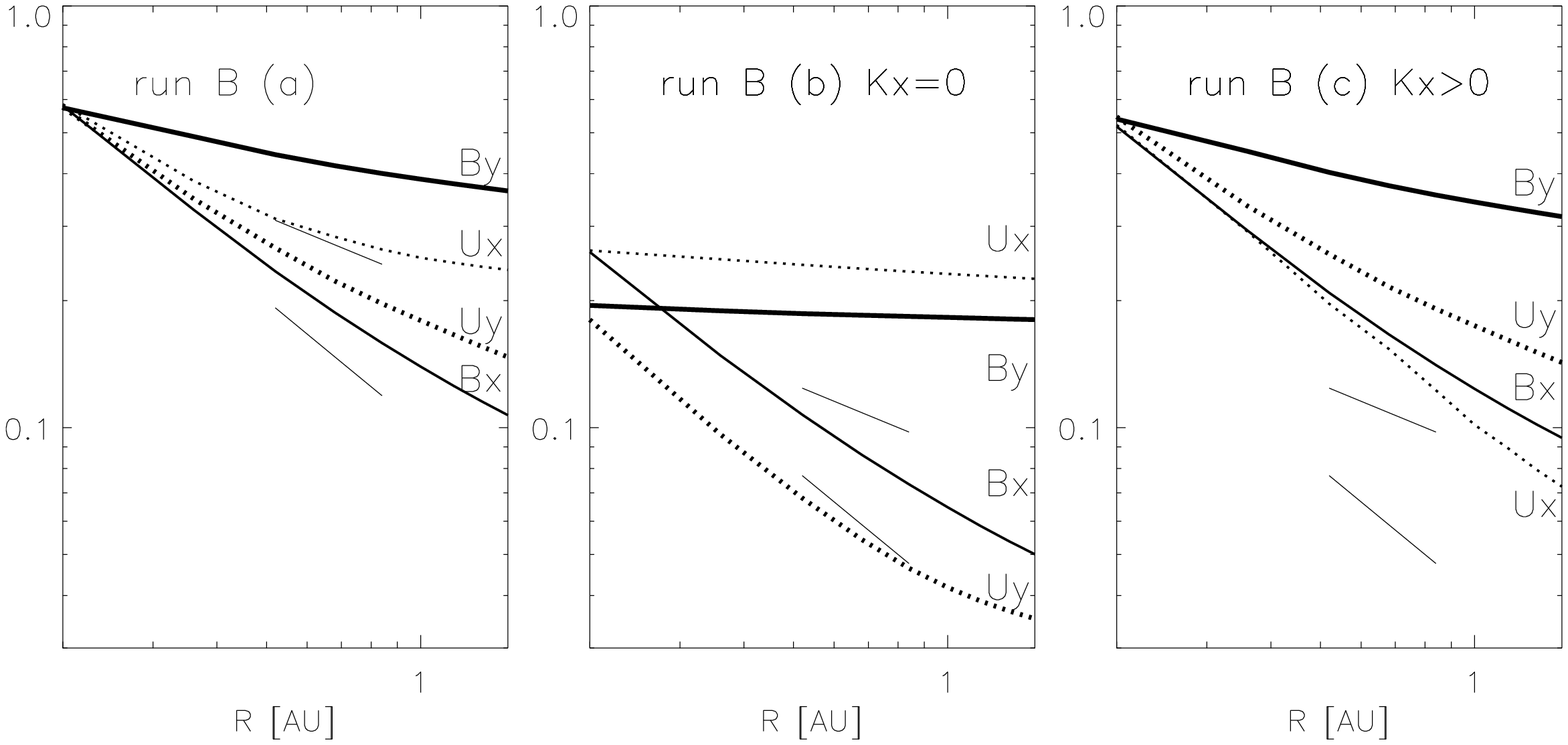}
\caption{
Run B. Decay of rms amplitudes with distance ($R$) for the different components of velocity and magnetic field in alfv\'en speed units. From left to right: (a) total rms
amplitudes, (b) rms amplitudes of the 2D modes ($K_x=0$, eq.~\ref{xls}), (c) rms amplitudes of the 3D modes ($K_x>0$, eq.~\ref{xss}). 
In all panels, the short solid lines indicate the $1/\sqrt R$ and $1/R$ scalings. 
Panel (b) shows that the decay of 2D modes is close to that imposed by the sole linear coupling with expansion, see text.
%Decay of rms amplitudes of different polarizations vs heliocentric distance $R$, run B.
%Note $B_x$ and $B_y$ are given in alfv\'en speed units, here and in
%the following.
%Left (a) : total rms amplitude, center (b) : $K_x=0$ contribution to the rms
%amplitude, right (c) : $K_x >0$ contribution to the rms amplitude. 
%Short solid lines: $1/\sqrt R$ and $1/R$ scalings. See definitions in text (eqs.~\ref{xls}-\ref{xss}).
}
\label{figpolarB}
\end{center}
\end{figure} 
\begin{figure*}[th]
\begin{center}
\includegraphics [width=0.45\linewidth]{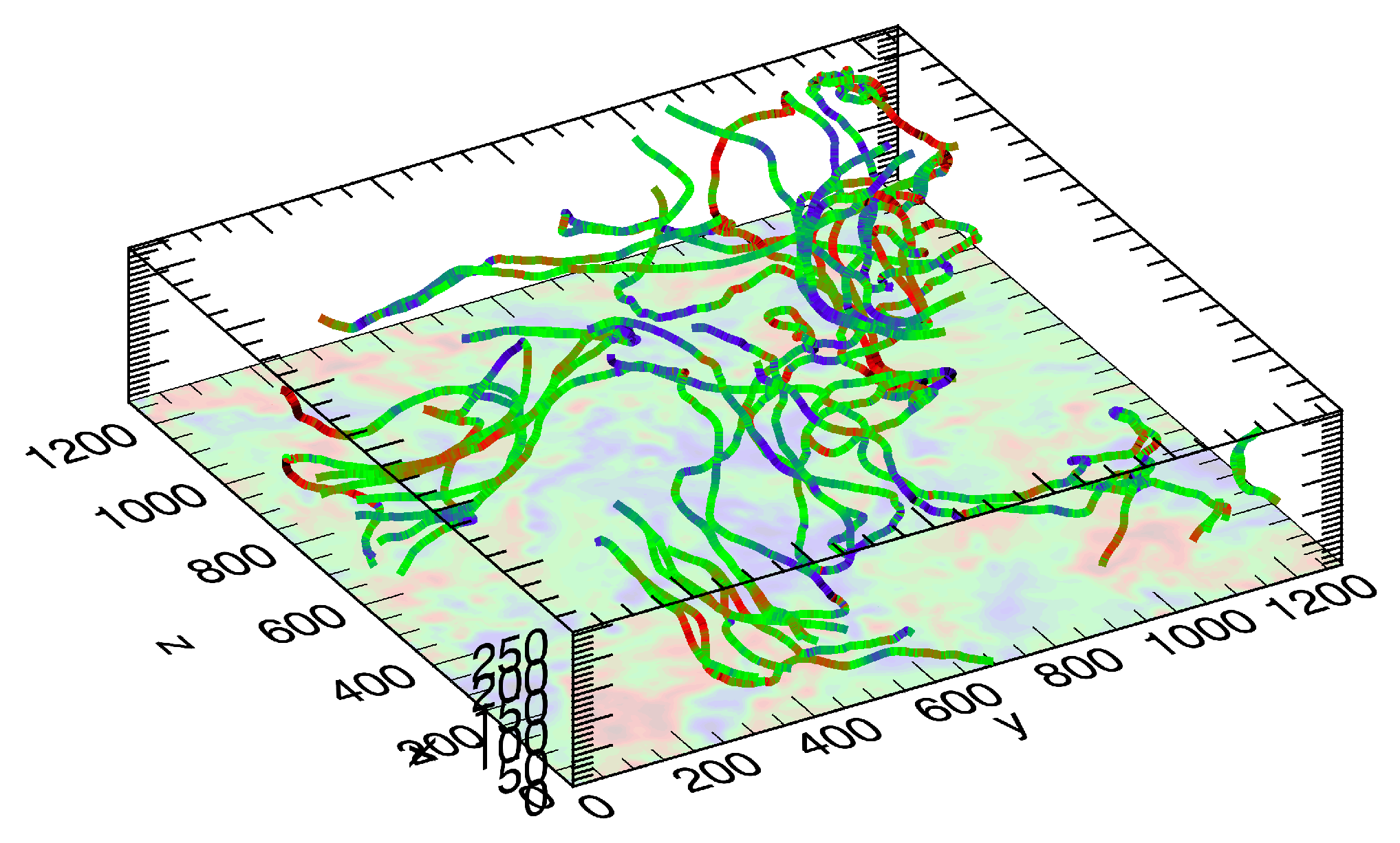} 
\includegraphics [width=0.45\linewidth]{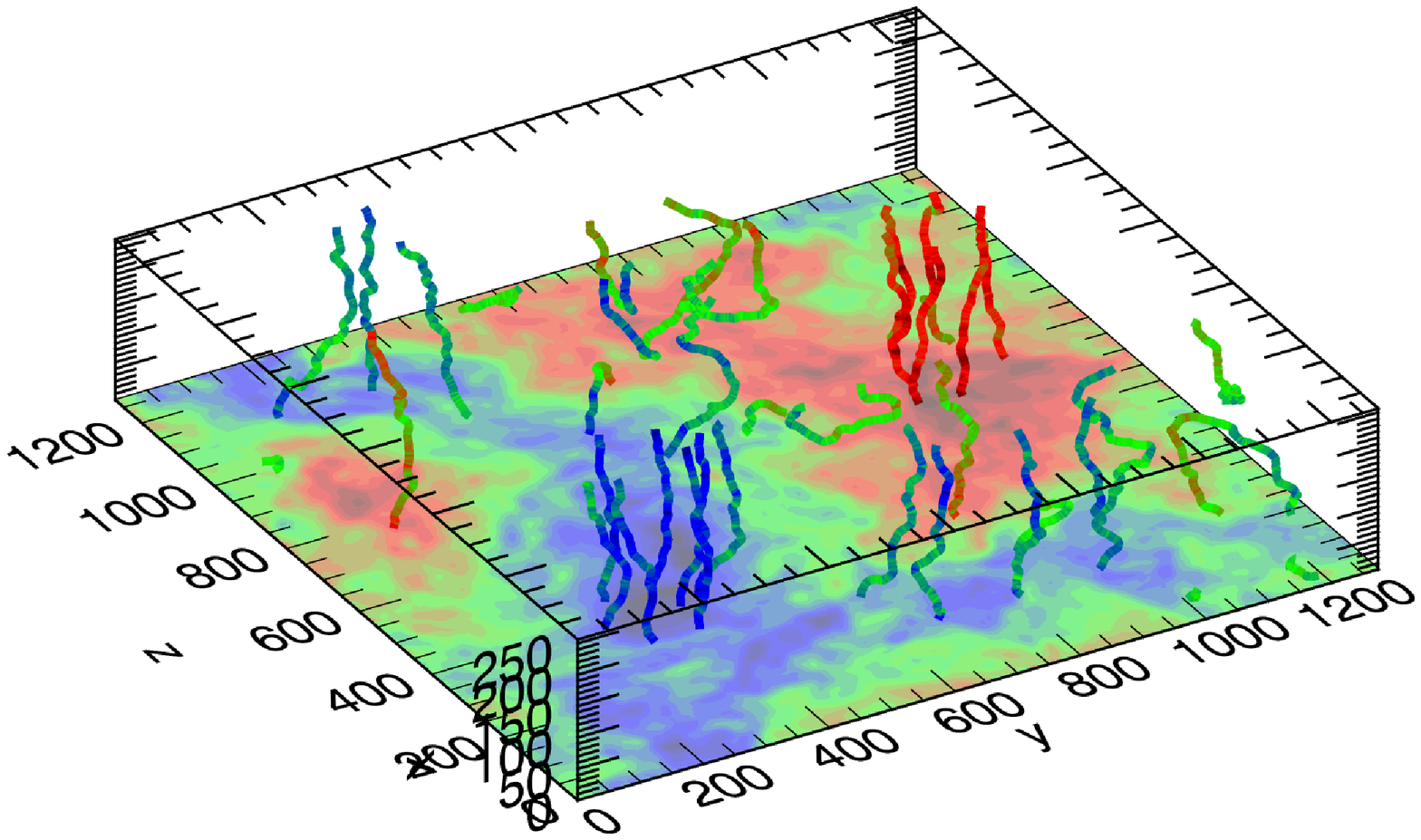}\\
\includegraphics [width=0.05\linewidth]{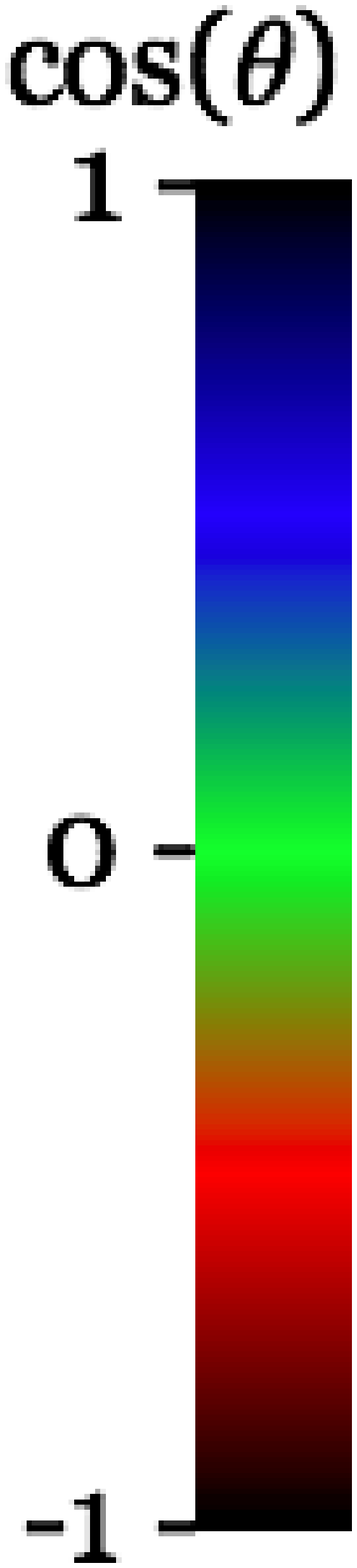}
\includegraphics [width=0.25\linewidth]{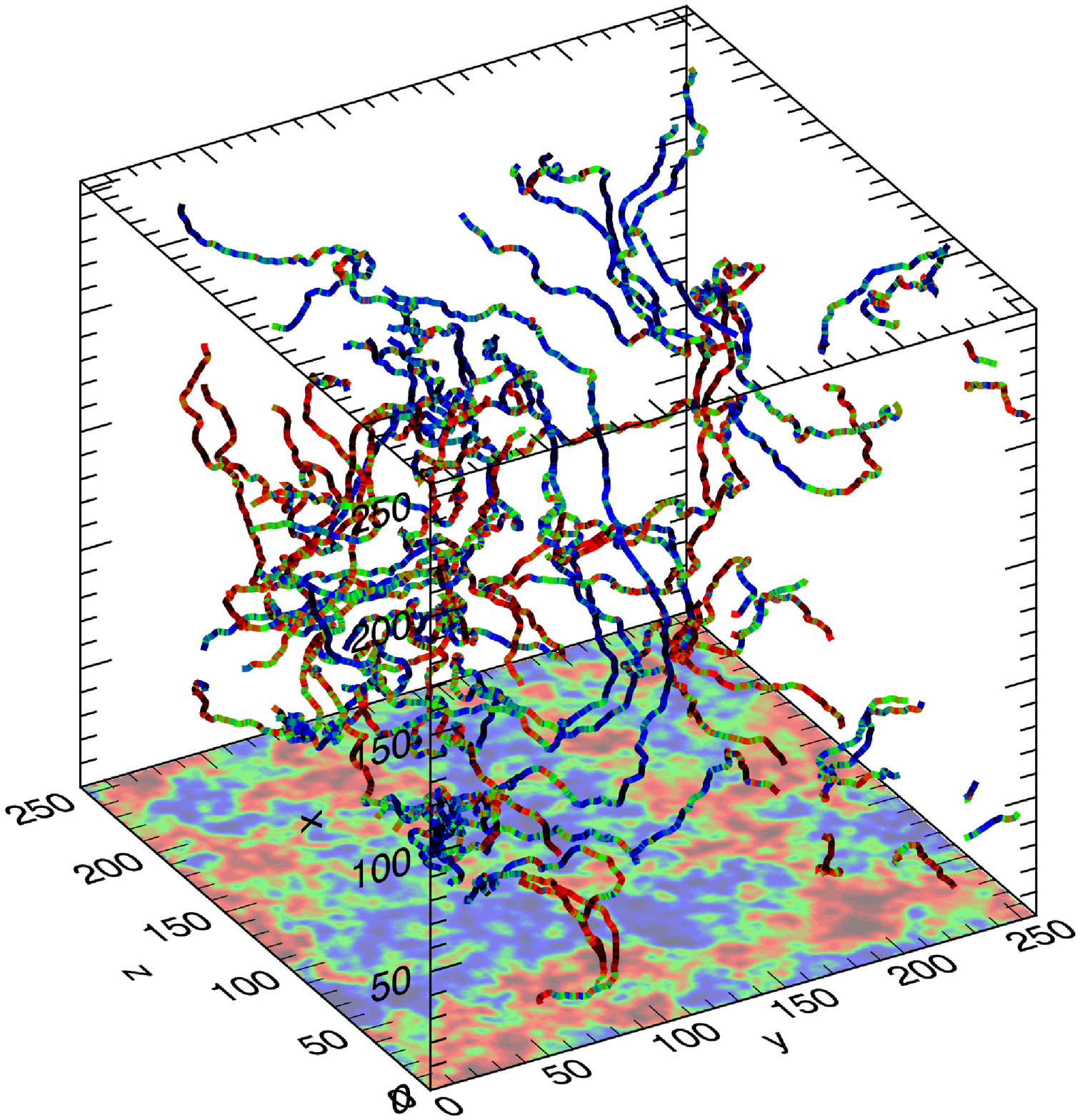} 
\includegraphics
[width=0.25\linewidth,trim=3.cm 13.cm 2.cm 2.cm, clip=true]{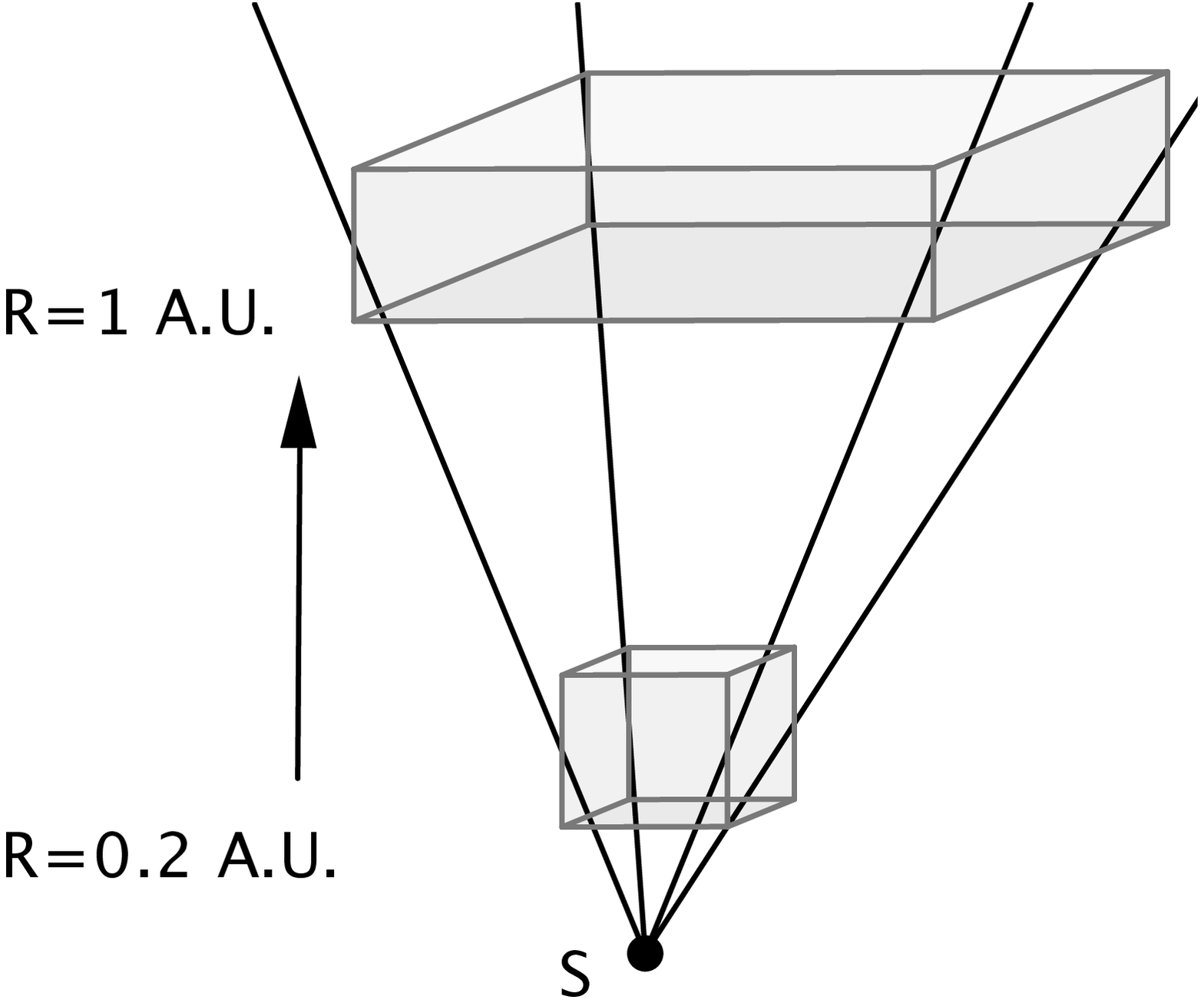}
\includegraphics [width=0.25\linewidth]{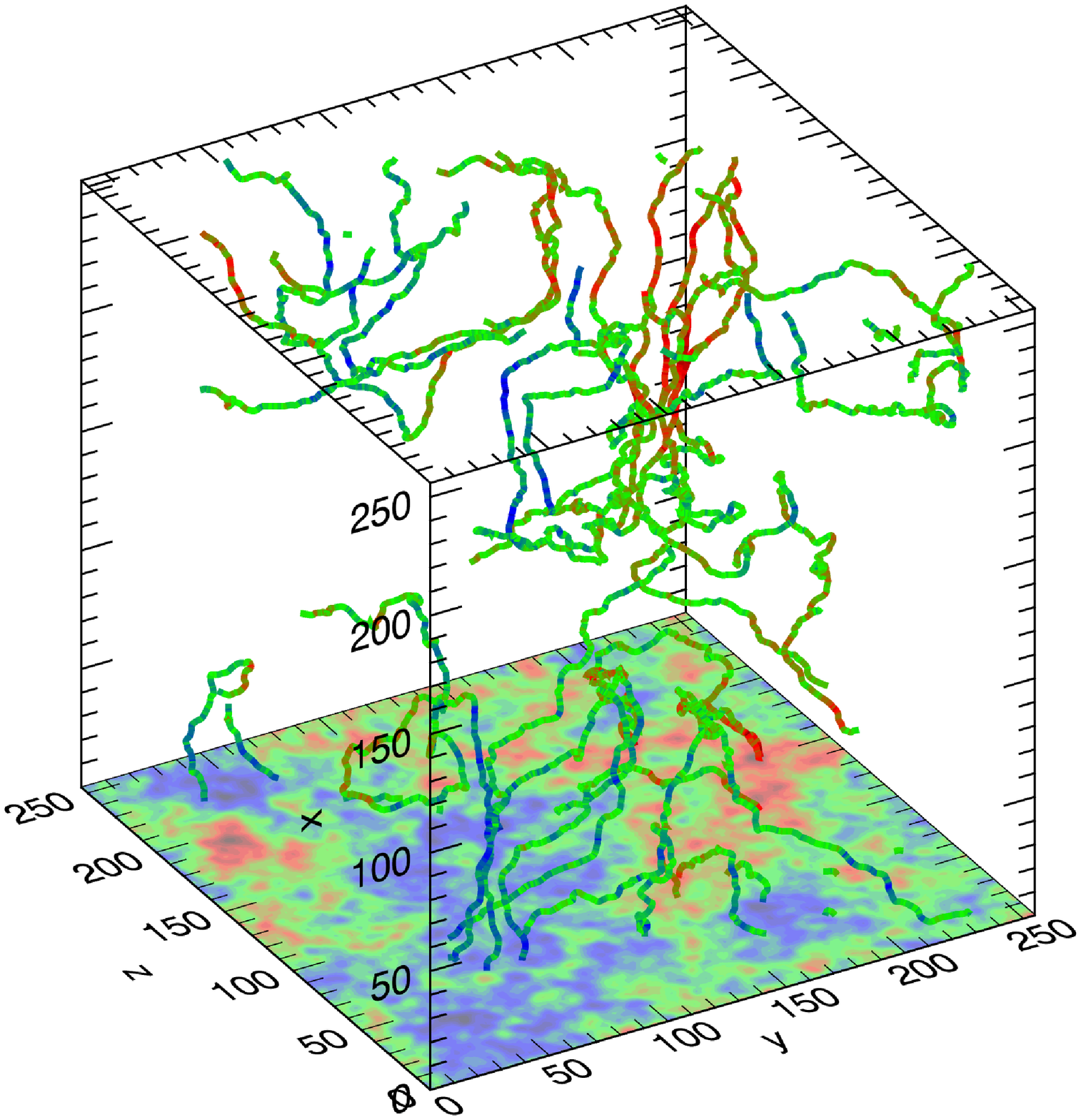}
\includegraphics [width=0.05\linewidth]{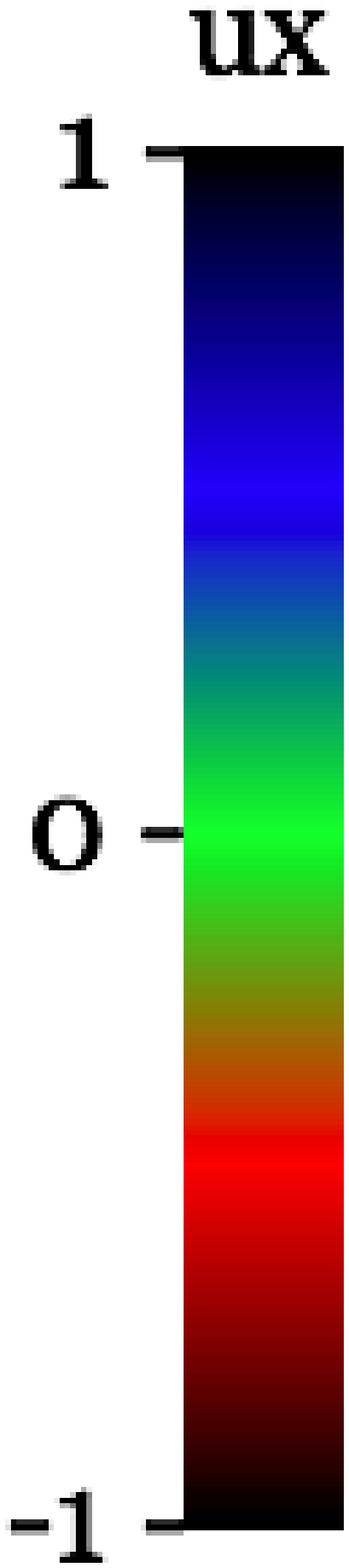}
%\includegraphics [width=0.49\linewidth]{HRES/b0e2_vecf_b_t2_rot0_ph_br3_eps.eps} 
%\includegraphics [width=0.49\linewidth]{HRES/b0e2_vecf_v_t2_rot0_6_eps.eps}
%\includegraphics [width=0.05\linewidth]{HRES/bx_table_eps.eps}
%\includegraphics [width=0.25\linewidth]{HRES/b0e2_vecf_b_t0_rot0_ph_br_eps.eps} 
%\includegraphics [width=0.25\linewidth]{HRES/expansion_dessin_box5_vert.eps} 
%\includegraphics [width=0.25\linewidth]{HRES/b0e2_vecf_v_t0_rot0_6_eps.eps} 
%\includegraphics [width=0.05\linewidth]{HRES/ux_table_eps.eps}
%\end{tabular}
\caption{
Expansion induced magnetic and kinetic fields anisotropy. Run B.
Left : magnetic field lines at 0.2 (initial condition, bottom) and 1 A.U. (top), colors
show the cosine of the angle of the magnetic field with radial
direction $\cos(\theta),\ \theta=(\vect{B},\vect{x})$. 
Right: velocity field lines, colors show the
amplitude of the radial velocity $U_x$ (normalized by the maximum). For each plot, contours show the same
quantity in the $x=0$ plane. Distance units are arbitrary, but takes into
account the aspect ratio due to expansion.
While the magnetic field lines tend to
align perpendicular to radial direction, velocity field lines tend to
create radial flux tubes.
}
\label{fig:first_overview}
\end{center}
\end{figure*}

To understand the effects of expansion on turbulence, we will start
by examining the radial evolution of scalar quantities in presence of
expansion. We will look at run B, which has some expansion ($\epsilon=2$) and no mean magnetic field. 

Figure~\ref{figpolarB}a shows the decay of the root-mean-squared (rms)
amplitude of radial and perpendicular
components, for both the velocity and magnetic fluctuations (in Alfv\'en unit).
At large distances the perpendicular magnetic component ($B_y$) and the
radial velocity component ($U_x$) dominate.
The latter actually decays faster than the former during an initial phase,
but finally both quantities decay at the same rate. 
The ordering of rms amplitudes in the different components is given by 
\begin{equation}
  \label{eq:liner_beha0}
 B_y>U_x>U_y>B_x
\end{equation}
This ordering may be interpreted as being due to the combined effects of expansion (eqs.~\ref{exp1}-\ref{exp4})
and nonlinear decay. 
A deeper understanding is obtained by considering separately the two components or the rms amplitudes, namely
the 2D modes with only wave vectors transverse to the radial direction ($K_x=0$), and the full 3D modes with
$K_x >0$:
\begin{align}
X_{2D}^{(\alpha)} = (2 E^{(\alpha)}(K_x=0))^{1/2}
\label{xls} \\
X_{3D}^{(\alpha)} = (2 \Sigma_{K_x > 0} E^{(\alpha)}(K_x))^{1/2}
\label{xss}
\end{align}
with the total rms amplitude verifying $X_{rms} = (X_{2D}^2+X_{3D}^2)^{1/2}$, and with
the index $\alpha$ denoting one of the field components ($U_{x,y}$ or
$B_{x,y}$). 

In the middle and right panels we draw separately the two rms amplitudes $X_{2D}$ and $X_{3D}$.
The panel (b) shows that the decay of the 2D modes is close to that imposed by the linear expansion (eqs.~\ref{exp1}-\ref{exp4}):
\begin{equation}
  \label{eq:liner_beha}
 B_y,~U_x \sim const > B_x,~ U_y \sim 1/R
\end{equation}
On the other hand, the decay of 3D modes (right panel) is affected by
expansion and nonlinear turbulent dynamics in a non trivial way, showing a
strong damping of radial velocity fluctuations ($U_x$) and yet another different ordering:
\begin{equation}
  \label{eq:liner_beha2}
 B_y >U_y> B_x> U_x
\end{equation}
Note that the $z$ component (not shown in the figure) behaves as the $y$ component, as expected since the $Ox$ axis is the symmetry axis.
The asymptotic behavior of rms amplitudes (left panel) is determined
by 3D modes, with the exception of the radial
velocity, $U_x$ (and perhaps $B_y$ at
larger distance) that is determined by the non-decaying 2D modes (central
panel) that are dominant.

The dominant degrees of freedom that emerge from our simulations are thus the radial component of the velocity $U_x$, that is quasi 2D, and the perpendicular components of the magnetic field $B_{y,z}$, that are fully 3D.
A snapshot of the velocity field and magnetic field immediately reveal these two structures.
In fig.~\ref{fig:first_overview} we represent the magnetic field
lines (left) and the velocity field lines (right) at two different times:
the initial time corresponding to $R= 0.2$ AU (bottom) and the final time
($t=2$) corresponding to $R= 1$ AU (top).
The colors indicate the cosine of the angle of the magnetic field with the radial ($B_x/|B|$, left) and the amplitude of the radial
velocity normalized to its maximum ($U_x/|U_x|_{max}$, right).

Comparing the bottom and top panels, one sees that the initial isotropy imposed at $0.2 AU$ has disappeared at $R= 1 AU$: the large scale velocity is now mainly made of radial streams, while the magnetic field is made of transverse lines.
Such a behavior has already been observed in the 2D simulations of the EBM by
\citet{Grappin_1996} and is a consequence of the differential radial decays of the components shown in the
previous figure. In summary:
\begin{enumerate}
\item Since $U_x$ and $B_y$ dominate over the other components they form
two kinds of visible spatial ``structures'', respectively radial velocity
streams and transverse magnetic field lines.
\item For $U_x$, the 3D ($K_x>0$) component falls down rapidly and reaches a
very low level, while the 2D ($K_x=0$) component remains almost constant.
The radial velocity streams will depend very weakly on the radial coordinate $x$, i.e. they appear uniform in the radial direction.
\item For $B_{y,z}$ the 3D component dominates over the 2D component, so
the magnetic field lines, mainly with transverse components, will appear more turbulent compared to velocity
field lines.
\end{enumerate}

A detailed examination of the figure shows that
the position of the velocity streams coincide with the position of random
maxima in the initial conditions. If we apply a low-pass filter to the
simulation, we find approximately the same $U_x(y,z)$ structures, in position
and amplitude from 0.2~AU (initial state) to 1~AU. 
Hence we can interpret the appearance of the structures as resulting from the \textit{selective decay} of some of the components. 
This spontaneous emergence of structures will occur as far as 
a reasonably isotropic distribution of energy among the different
components is present in the initial conditions close to the Sun.
Note that the original selective decay idea was devised for flows
decaying due to turbulent viscosity by \citet{1978PhFl...21..757M}
while here the decay (and thus the appearance of the stream and
magnetic structures) is due to the sole wind expansion.

%------------------------------------------------------------
\subsection{Turbulent energy decay and spectral formation}
We first consider the decay of total energy with time (distance) in four runs
A,B,C, and D, then we analyze their spectral evolution. Finally we describe
in detail their
spectral and component anisotropy at a given time (corresponding to a
heliocentric distance of about 1~AU).
\subsubsection{Energy decay}
A main feature of (unforced) turbulent flows is turbulent energy decay. 
In a standard simulation at small Mach number (as here) of a decaying flow from a large scale reservoir, one expects that during a finite time of order of
the nonlinear time the total energy will be conserved as in an incompressible
flow (e.g., \citealt{2010GApFD.104..115P}). 
This initial phase should be followed by a phase of fast decay, which overruns by several orders of magnitude the ordinary, laminar dissipation in an ordinary fluid.

Since we are considering decaying turbulence, we expect the same behavior, however
with several important differences.
First, small scales are present from start in our simulations, since the initial spectrum scales as $K^{-1}$.
We may thus expect turbulent energy dissipation to start earlier than in the case of an initial spectrum concentrated at large scales.
Second, in the case of expanding runs ($\epsilon\ne0$), 
the conservation of quadratic invariants is replaced by the
conservation of first-order invariants (see Section~\ref{LEE}). Thus, the kinetic and magnetic energies
will be strongly damped even in the absence of turbulent dissipation.
\begin{figure}
\begin{center}
\includegraphics [width=\linewidth]{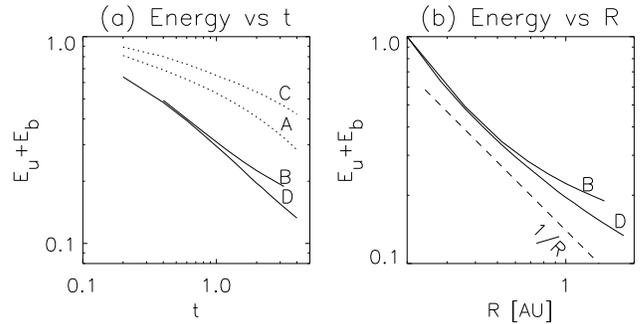}
\caption{
Total energy (sum of kinetic and magnetic energies) decay
(a) Energy for runs A, B, C, and D as a function of
time $t$ in units of nonlinear time 
(b) Energy for runs B and D with expansion as a function of distance $R$.
The dashed line indicates the $1/R$ wkb energy decay law as in eq.~\eqref{wkb}.
}
\label{fig0}
\end{center}
\end{figure}

The energy (kinetic + magnetic) decay with time is shown in Figure~\ref{fig0}-a for the first four runs of Table~1. Time is normalized by the initial nonlinear time.
One sees indeed that the energy decay begins from start in all four cases, but
at a much slower rate compared to the Navier-Stokes case in which $E \sim
t^{-10/7}$ (e.g., \citealt{1995tlan.book.....F}). 
For the ``expanding'' runs the energy decays at most as $E \sim t^{-1/2}$, and significantly slower for the runs without expansion.
A possible cause of such a slow decay lies in the initial $k^{-1}$ spectrum that
progressively transforms into a steeper spectrum (as we will see in the next
section). Correspondingly, the energy containing scale that feeds the cascade increases with time, so leading to the observed slow-down of the energy decay. 
As a rule, expanding runs (thick lines) decay faster than their
non-expanding counterpart (thin dotted lines). 
Also, a striking point is that, while in the homogeneous case the presence of a mean field slows down the decay, the contrary is true when expansion is present: the mean field then accelerates the energy decay
, instead of slowing it down.

In fig.~\ref{fig0}-b we examine the energy decay with heliocentric distance $R$
for the two ``expanding'' runs, the dashed line gives the $1/R$ WKB decay as a reference. While at the very beginning of the evolution the decay is faster than WKB, the reverse is true for distances $R/R_0 \ge 3$.
These different energy evolutions are related to the previous
component anisotropy but also to different spectral evolutions, which we examine now.

\subsubsection{Spectral evolution}
\begin{figure}
\begin{center}
\includegraphics [width=\linewidth]{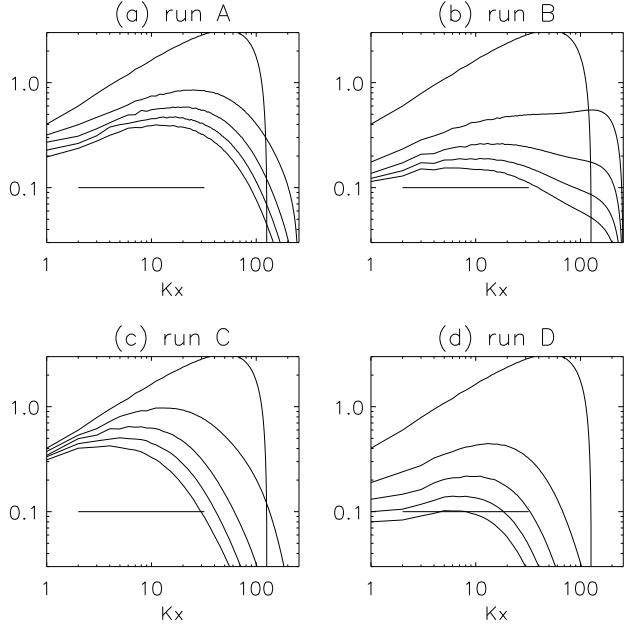}
\caption{
Time evolution of 1D reduced energy spectra (kinetic + magnetic energy) vs $K_x$ wavenumber.
All spectra are compensated by $K^{-5/3}$.
Runs A, B, C, D. Times shown are $t=0,~0.8,~1.6,~2.4,~3.2$
(unit time = initial nonlinear time). In the ``expanding'' runs B and D, the
corresponding distances are $R/R_0=1,~2.6,~4.2,~5.8,~7.4$, respectively.
}
\label{fig-spred}
\end{center}
\end{figure}

We show in Fig.~\ref{fig-spred} (left column) 1D reduced spectra vs radial
wavenumber $K_x$ for times $t=0,~0.8,~1.6,~2.4,~3.2$ (upper to lower curves
respectively), for the same four runs. All spectra are compensated by the $k^{-5/3}$ scaling.

In all four runs, the initial spectrum follows a $k^{-1}$ scaling, which is terminated by a sharp cutoff due to a spherical (isotropic) truncation at $k =128$.
Although the evolution is significantly different for the four runs,
all of them share the following properties at the last time $t=2$ shown in the figure:
(i) a large scale range with a spectral slope in between the initial slope $-1$ and the slope $-5/3$
(ii) a medium scale range showing a $k^{-5/3}$ scaling
(iii) a small scale dissipative range.

Two important remarks are in order.
First, the persistence of a relatively flat large-scale range is not a property specific of the expanding runs.
The origin of the persistence is thus to be found not only in the specific freezing of large scales due to expansion.
Second, the extent of the $k^{-5/3}$ scaling, which is a priori a signature of the nonlinear cascade is not more extended in the homogeneous than in the expanding runs. The contrary seems to be true for the zero mean field case (compare panels a and b): this could be the consequence of a higher average Reynolds number in the case of run B, due to the $1/R$ variation adopted for the viscosity.

%The reason is probably due to a strong reduction of the effective Reyonolds
%number in this particular configuration with radial mean magnetic field. If on the one hand, the presence of a mean field favors the cascades towards smaller
%field-perpendicular scales, on the other hand the expansion of scales
%perpendicular to the radial induces a cinematic decrease of $Re_{eff}$. We examine these anisotropic effects now.
%
%The reason is
%probably that, in the presence of a mean field, the energy cascade to small
%scales mainly in the plane perpendicular to the mean-field axis, which here coincides with the plane perpendicular to the radial. But since the directions perpendicular to the radial are expanding, this should induce a decrease of the $Re_{eff}$. We examine these anisotropic effects now.
%

%------------------------------------------------------------

\subsubsection{Spectral (gyrotropic) anisotropy}
We consider now the energy distribution of total energy among wavevectors that allows us to visualize the effect of expansion and that of a mean field on the spectral anisotropy.
%We consider now the spectral anisotropy that results either from the presence
%of the expansion or from that of a mean field. 
In the run with radial mean field examined here (runs C and D), we have checked that the energy spectra are reasonably gyrotropic around the x axis.
We will consider later in the Discussion deviations from gyrotropy, 
when we come to examine run E with a non radial magnetic field.

\begin{figure}[t]
\begin{center}
\includegraphics [width=0.45\linewidth]{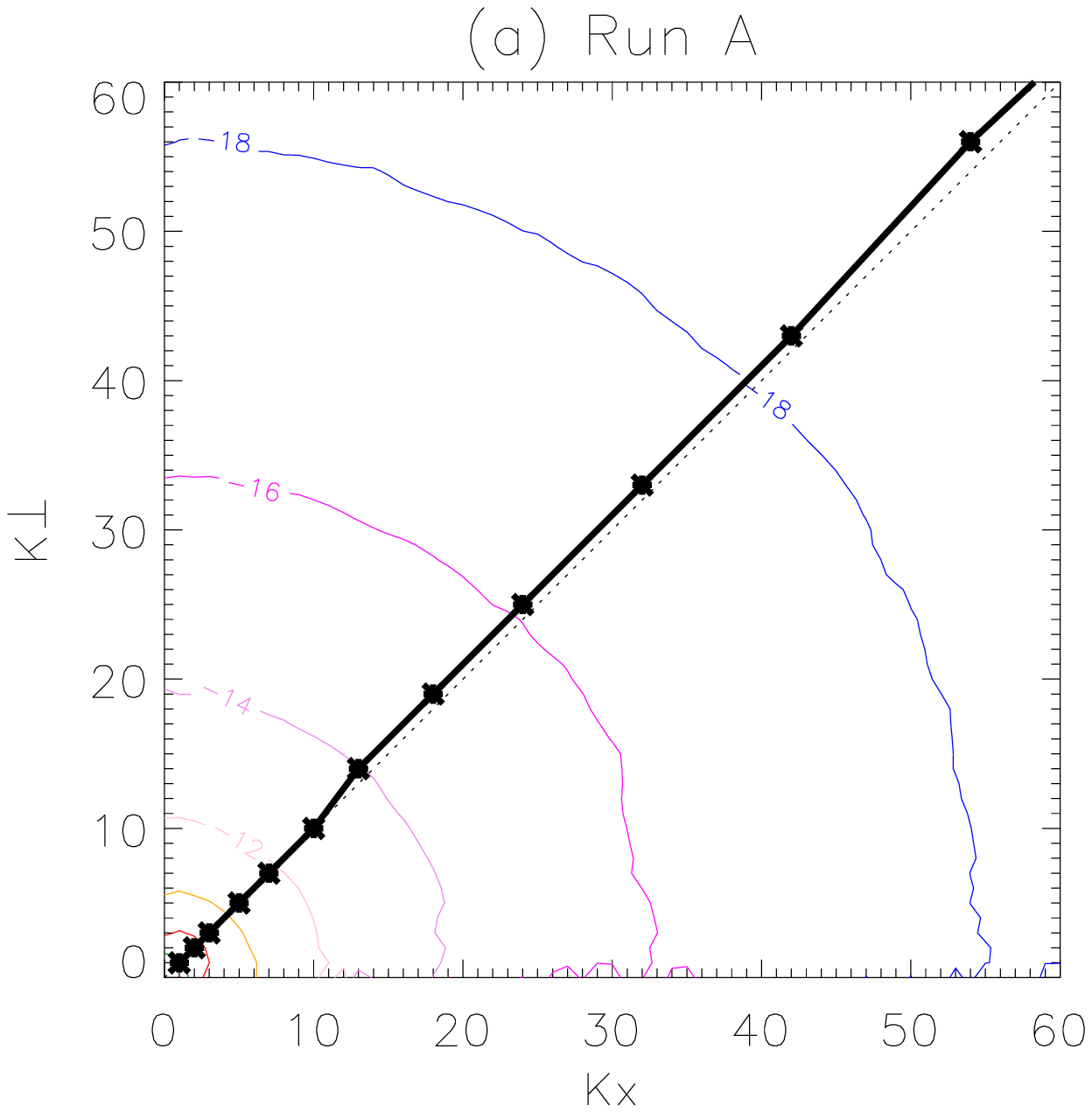}
\includegraphics [width=0.45\linewidth]{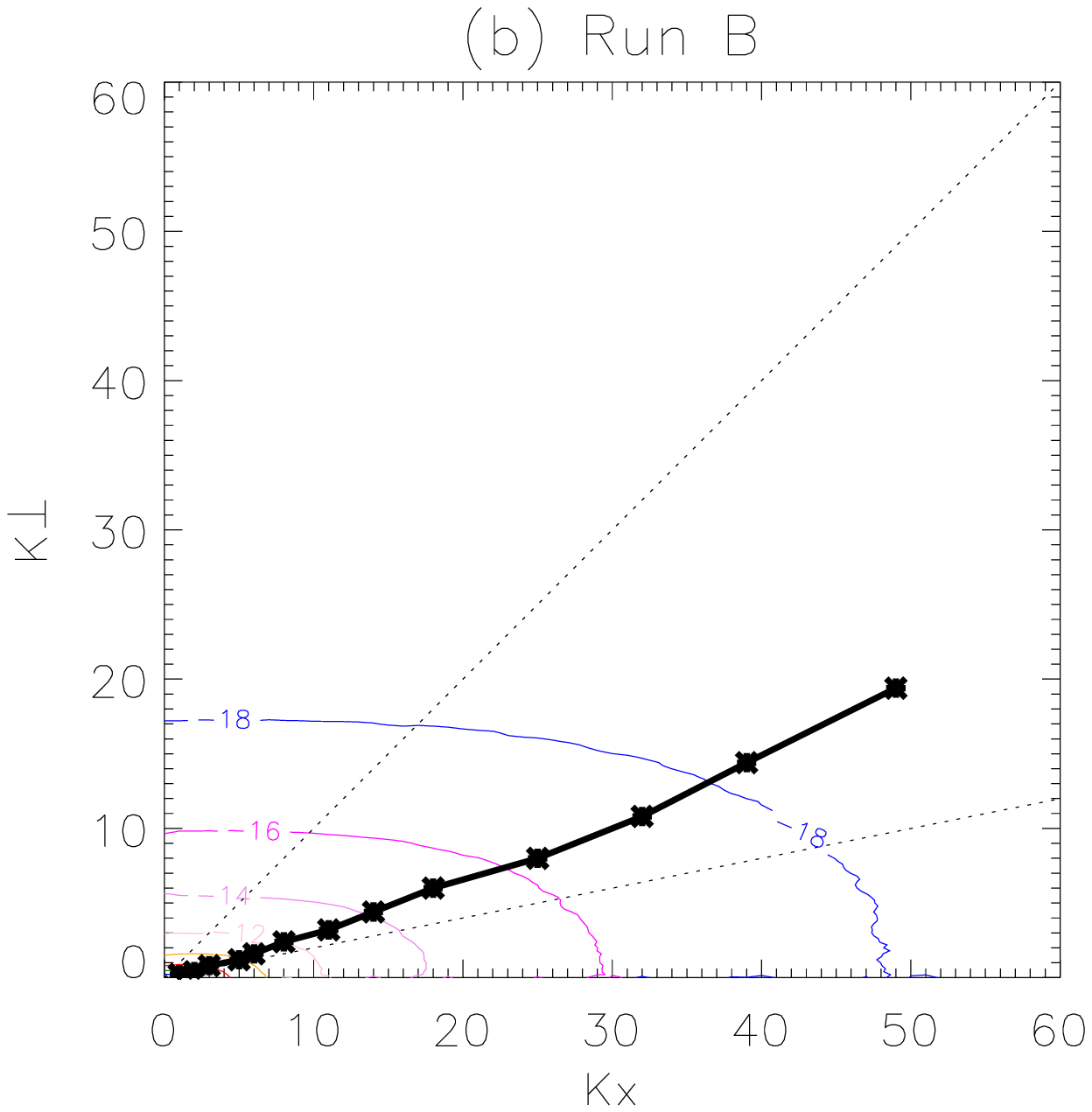}
\includegraphics [width=0.45\linewidth]{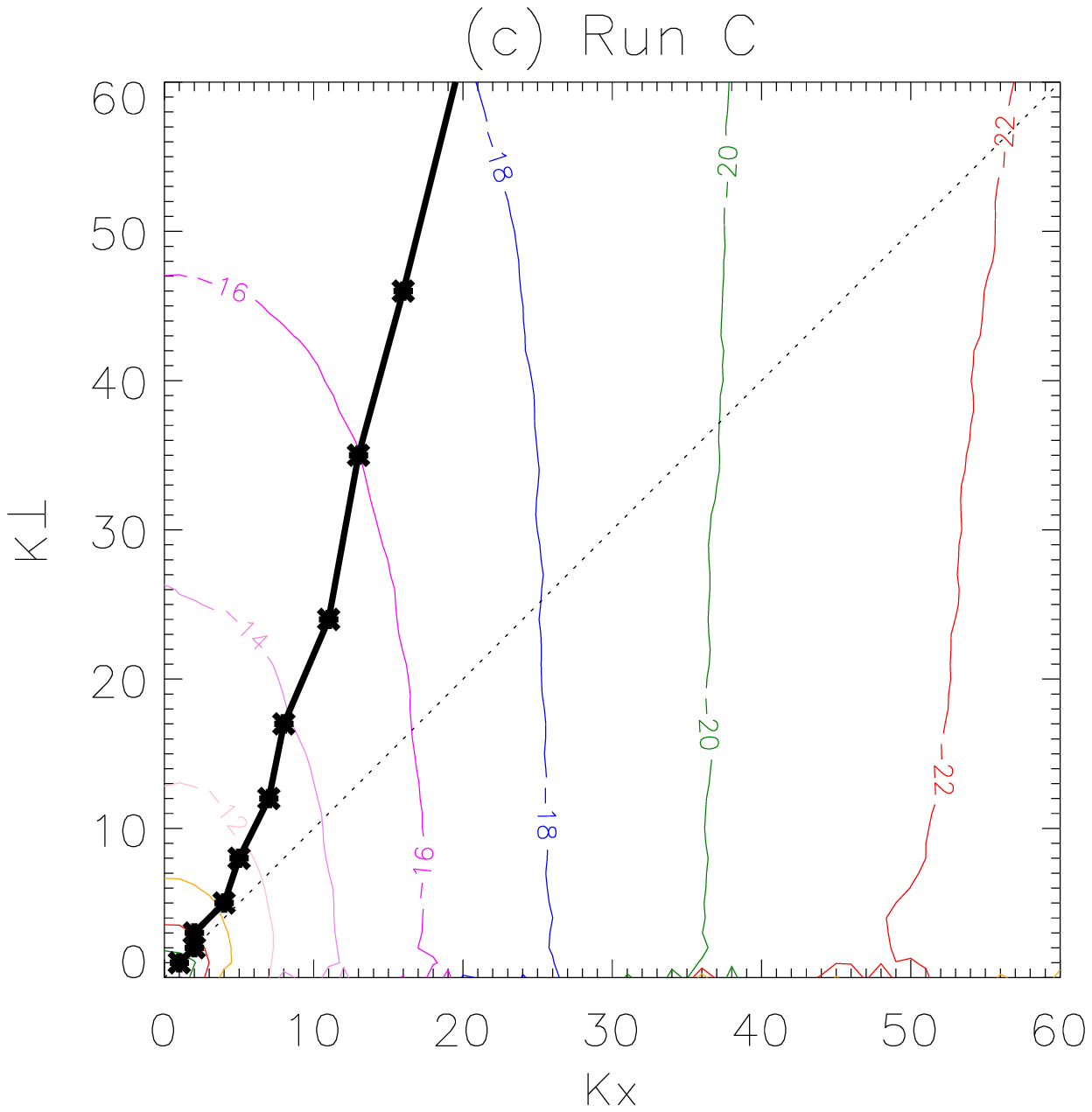}
\includegraphics [width=0.45\linewidth]{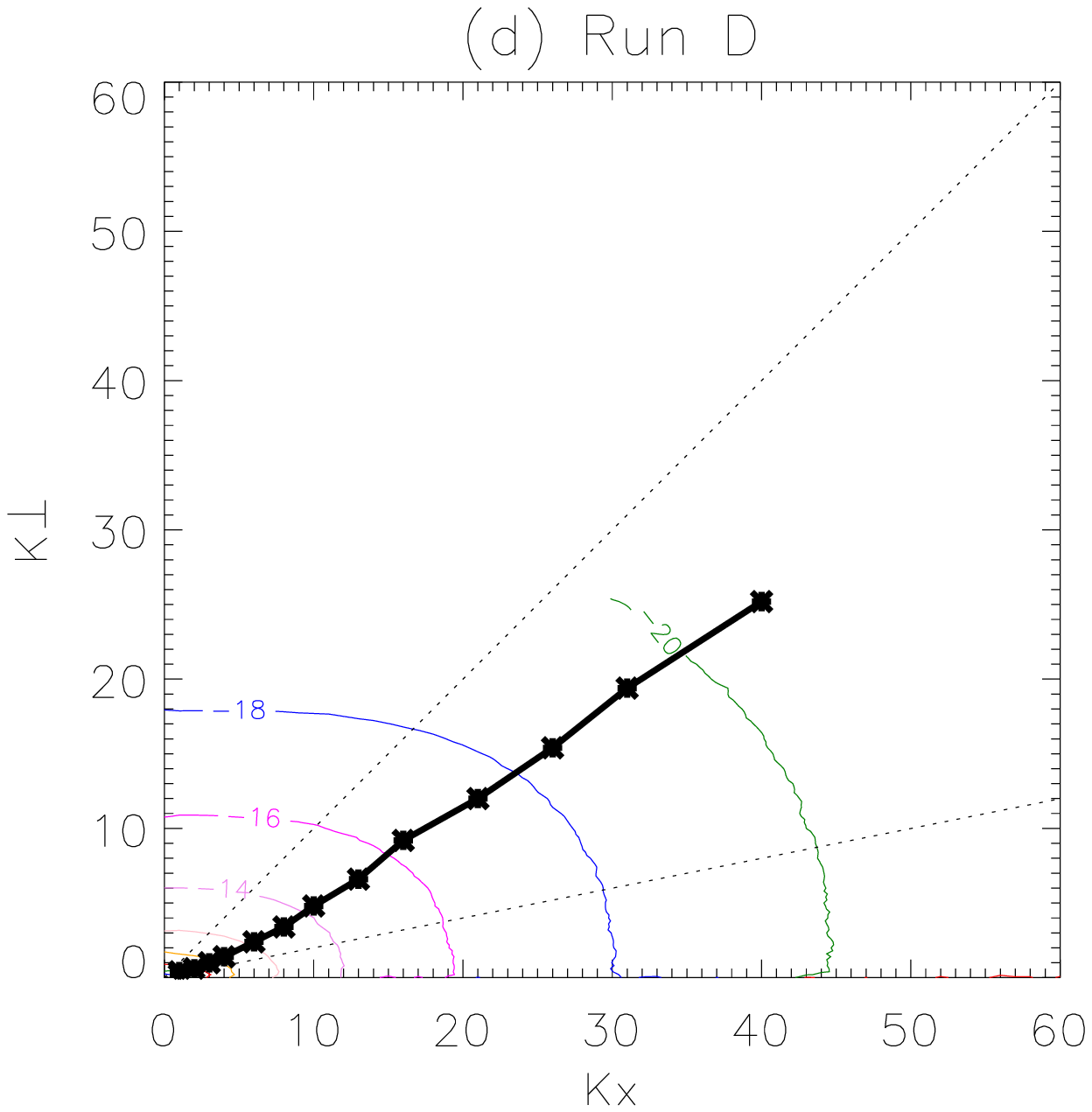}

%----
\caption{Spectral anisotropy.
Isocontours of the Magnetic energy spectrum ($E_{3D}^{gyro}$) in
\textit{physical} wavenumbers ($K_x, K_\bot$), at time $t=2$.
From top to bottom:
(a) run C, mean field; (b) run B, expansion; (c) run D, expansion and mean
field.  
The thick line starting at the origin shows a measure of the anisotropy of the
spectrum (see text).
The kinematic anisotropy that would result from the sole expansion effect, reflecting the
aspect ratio of the plasma volume, is indicated in panels (b) and (d) by a dotted line.
}
\label{fig4}
\end{center}
\end{figure}

In fig.~\ref{fig4} we plot in the plane $(K_x,K_\bot)$ the isocontours of the magnetic 3D energy spectrum $E_{3D}^{gyro}$, averaged around the $K_x$ axis,  
for the four runs A, B, C and D at time $t=2$.
The anisotropy is visible in the aspect ratio of the
isocontours that show a systematic elongation either in favor of the $K_x$ axis or the $K_y$ axis, depending on the run.
To quantify the deviations from isotropy at different scales, we provide a
simple measure of the aspect ratio, by computing the intersections of
isocontours with the $K_x$ and $K_\bot$ axis and by plotting each couple of
$(K_x,K_\bot)$ points satisfying to $E(k_x,0) = E(0,k_\perp)$. 
These points are connected by a thick line in each panel, indicating which
direction, whether radial or transverse, is the most excited at each wavenumber
$|K|$. If the spectrum were fully isotropic, one would obtain the dotted
diagonal line, $K_x=K_\perp$, plotted in all the figures.
If all wavenumbers were frozen in (no nonlinear transfer), expansion would
lead for runs B and D to a collapse of isocontours in Fourier space on the parallel wavenumber axis $K_x$.
Their aspect ratio would correspond to the lower dotted-line (off the diagonal) in
panels (c) and (d). 

In panel (a) (run A, no expansion, no mean field), 
the spectrum is fully isotropic, the thick line is a diagonal.
In panel (b) (run B, expansion, no mean field)
the thick line indicates that the spectrum is more developed along the radial,
as expected in view of the perpendicular expansion.
%The anisotropy indicated by the inclination of the anisotropy line can be estimated as $B_0/b_{rms} = 2/0.75 = 2.7$ where $\tan(\theta)=100/35=2.8$.
In panel (c) (run C, no expansion, mean field), the very large-scales are isotropic (as shown by the thick line that follows the diagonal for $K<5$),
while higher wavenumbers depart strongly from the diagonal, showing 
the dominance of the energy flux in directions perpendicular to the mean field, as expected.
Finally, in panel (d), (run D, expansion and radial mean magnetic field), 
the anisotropy results from
to the competition between the opposite effects of the mean field and expansion on the cascade,
producing a weak spectral anisotropy along the radial direction.
At the very large scales ($K < 5$), the anisotropy follows the sole expansion effect.

%Quasi-Isotropy is reached because the mean field on the one hand
%favors the perpendicular cascade that is frozen at early time and follows the
%geometric deformation due to expansion, and on the other hand reduce the
%radial cascade, that is not affected by expansion.

The following approximate expression for the anisotropy profile $\cal A(K)$ puts together in the simplest possible way the basic bricks of the competition between expansion and perpendicular cascade:
\begin{equation}
  \label{eq:defA}
  {\cal A}=(B_0+b_{rms})/b_{rms})/(R/R_0)
%  \label{anisotropieA}
\end{equation}
One can check that it covers reasonably well all the four cases A, B, C and D considered up to now, with or without expansion, with or without radial mean field.

%\subsubsection{Deviation from equipartition}
\subsubsection{Component anisotropy}
\label{sec:devi-from-equip}

\begin{figure}[t]
\begin{center}
\includegraphics [width=\linewidth]{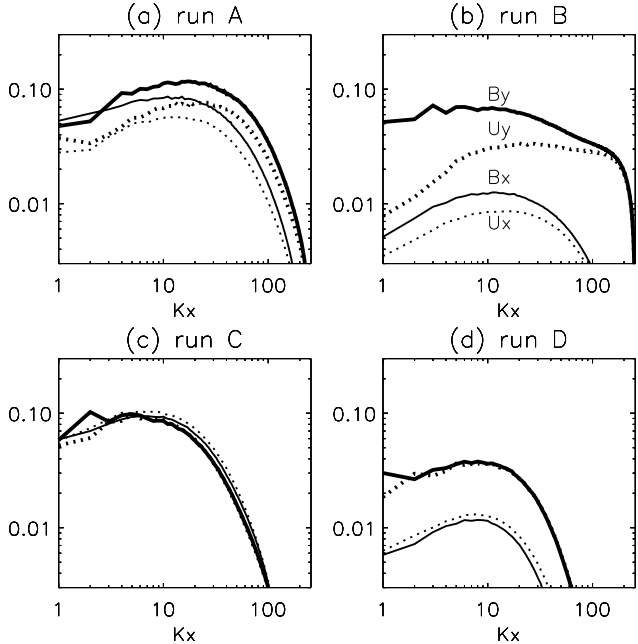}
%---------
\caption{
1D reduced spectra $E(K_x)$ compensated by $K^{-5/3}$ vs radial wavenumber $K_x$, for the different
components (see top right panel for styles) with magnetic field in Alfv\'en
speed units, for runs A, B, C, and D at time t=2.}
\label{fig2}
\end{center}
\end{figure}

Departure from energy equipartition between different components is clearly observed in solar wind turbulence,
so it deserves to be considered here (comparison will be made with observations in the discussion section).

We show in fig.~\ref{fig2} the 1D reduced spectra vs $K_x$ for the
four runs $A, B, C, D$ at $t=2$. We focus here on the ordering
of the different components of magnetic and velocity fields, rather
than their spectral index as previously. 
Except for run C which shows equipartition between all degrees of freedom, these orderings may be classified in two categories:
(i) a tendency to the dominance, component by component, of the magnetic field over the velocity field 
(ii) a dominance of the perpendicular components (here y) over the x (radial) component.

We start from the two homogeneous runs A and C, that in principle have known features.
The zero mean field case, run A, leads to a slight dominance of the y component, and to a magnetic excess, that reduces progressively to zero in the dissipative tail of the spectrum. This is valid for each component, both x and y.
First, we note that the observed slight dominance of the y components over the x components is the plain algebraic consequence of dealing with a quasi-incompressible flow. 
This is checked by considering $E(K_y)$ spectra (not shown): a similar (but reverse) dissymetry between the x and y components is observed.
Run C, with mean field case, shows on the contrary equipartition of all degrees of freedom at all scales.

In \citep{1983A&A...126...51G,Muller:2005jp}, a mechanism has been proposed that leads to such an excess, based on the competition between, on the one hand, nonlinear stretching of the magnetic field (called in the following ``local dynamo'') that systematically transfers energy from kinetic to magnetic and, on the other hand, the Alfv\'en effect (propagation along the local mean field) that leads to equipartition between the two fields.
The resulting magnetic excess is much larger in the zero mean field case, due to the reduced efficiency of the Alfv\'en effect compared to the non zero mean field case. In the mean field case, the difference between the magnetic and kinetic spectra actually fluctuates around zero from time to time, and a definite magnetic excess emerges only after averaging in time, which is not done here.

The cases of the two runs B and D with expansion are easily summarized.
Run B with zero mean field shows a much enhanced magnetic excess, component per component, and
the dominance of the y component can clearly not be attributed to the effect of quasi-incompressibility of the low, but must be a genuine effect of the expanding turbulence.
Run D with non zero mean field still doesn't show any measurable magnetic dominance, but instead shows a large dominance of the perpendicular component, absent in the corresponding homogeneous run C.

We propose below in the discussion an extension of the Alfv\'en-dynamo mechanism that includes expansion to explain these properties, and compare them with the ones observed in the solar wind.

\section{Discussion}

%------------------------------------------------------------
\begin{figure*}[t]
%\centering
\includegraphics [width=0.33\linewidth]{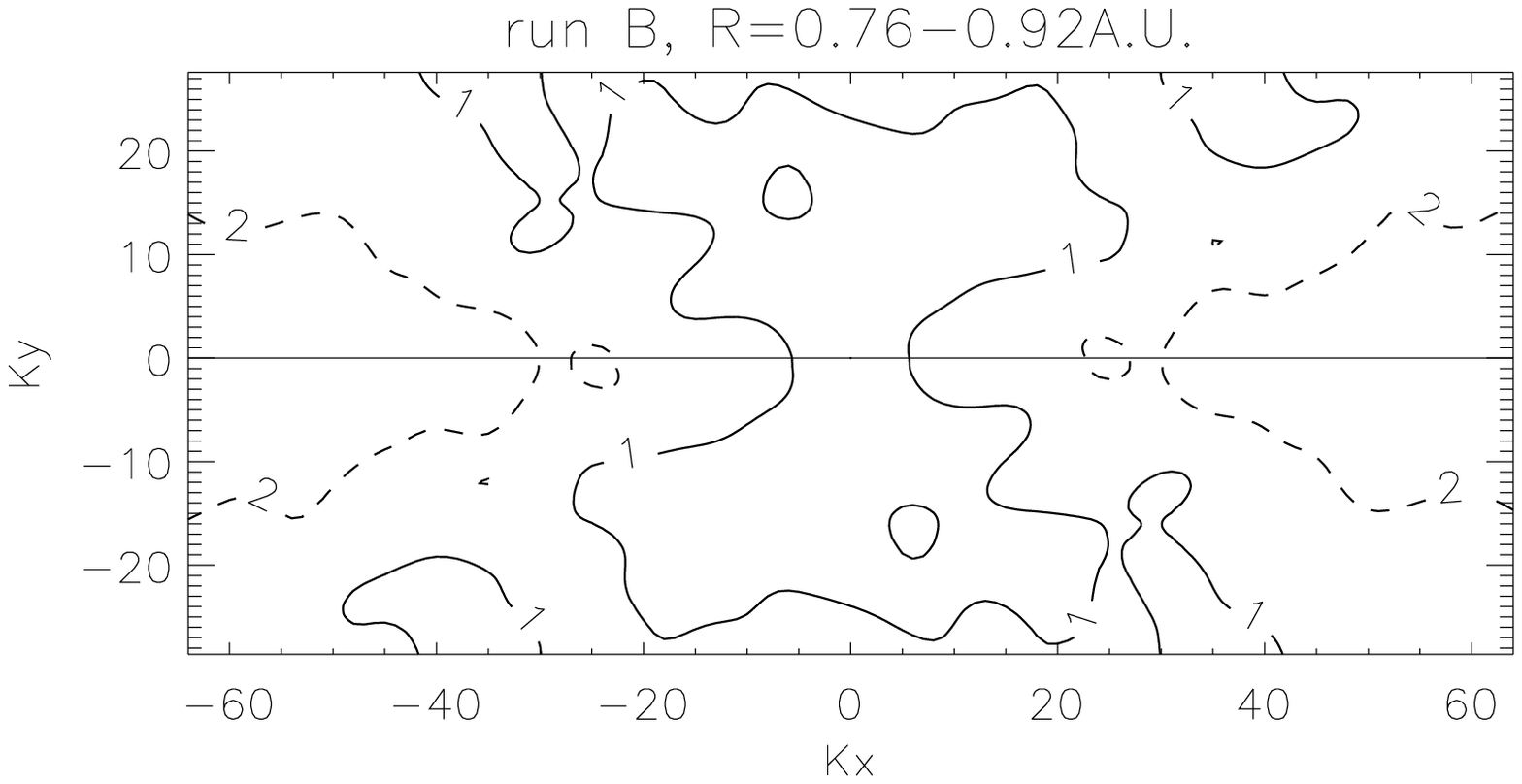}
\includegraphics [width=0.33\linewidth]{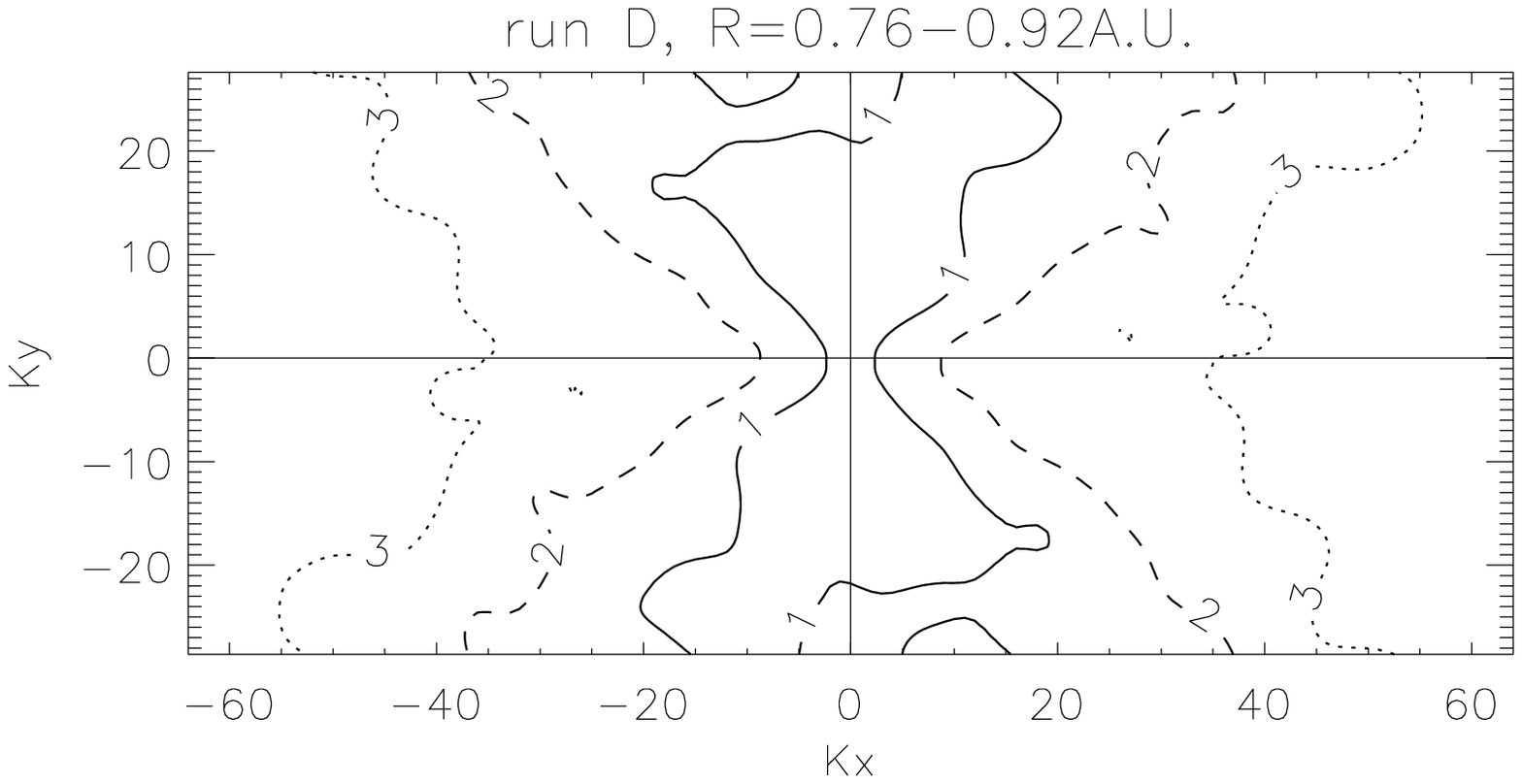}
\includegraphics [width=0.33\linewidth]{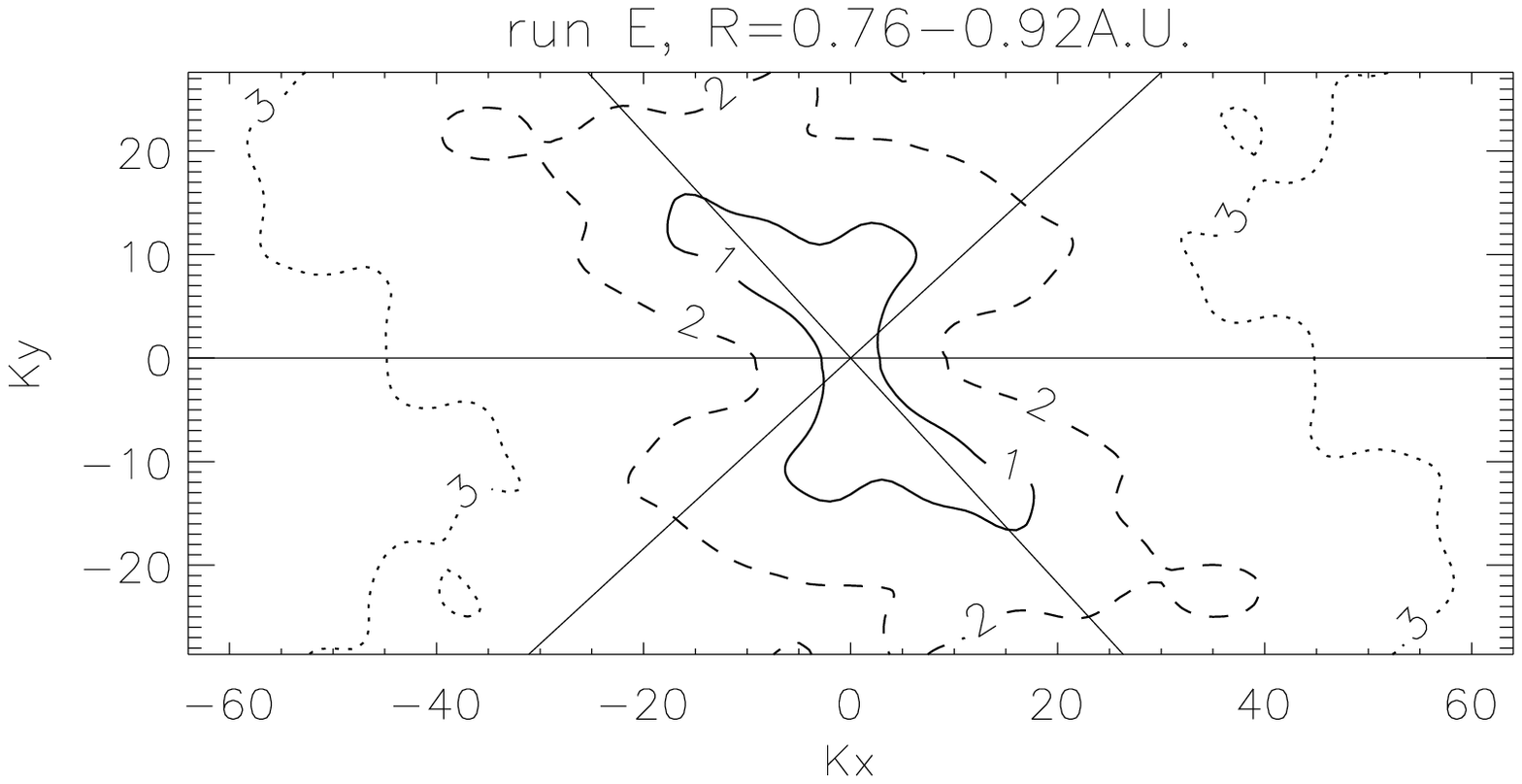}
\caption{
Runs B, D and E.
Visualizing the radial energy damping rate in Fourier space.
Lines: isocontours of \textit{radial damping rate} of total energy
spectra in the $(K_x,K_y,K_z=0)$ plane.
The radial damping rate $\alpha$ is given by $E(t_1)/E(t_0) =
(R(t_1)/R(t_0))^{-\alpha}$, with $t_0=1.4$, $t_1=1.8$ and
 $R(t_0)=0.76, R(t_1)=0.92$.
 (represented with vertical dotted lines in fig. \ref{fig7}).
Left panel, run B (no mean magnetic field); middle panel, run D (radial
mean magnetic field); right panel, run E (oblique mean magnetic
field). Thin lines indicate the directions parallel and perpendicular to the
mean magnetic field when present.
}
\label{fig8}
\end{figure*}
In this section we come back to several important points, focusing only on the
``expanding'' runs. We either discuss some remining issues or generalize some
of the results.

\subsection{Coherent structures: the case of mean field}
We have shown that coherent, stable, microjets-like structures emerge
spontaneously from the dynamics of turbulence with \textit{no mean-field} and
subject to \textit{solar wind expansion} (run B).
We further showed that the phenomenon is caused by the (linear) selective decay
of the large-scale fluctuations. The non-WKB damping (expansion) affects the
energy injection at large scales and the
resulting turbulent dynamics, eliminating some degrees of freedom
and leaving undamped some others ($U_x(K_x=0)$, $B_y(K_x=0),~B_z(K_x=0)$).
Note that a non-WKB damping implies that the Alfv\'en effect is not efficient in
coupling $U$ and $B$ component along $x$, casting some doubts on the
emergence of structures when a mean magnetic field is present. 
In fact, in this case one expects an efficient Alfv\'enic coupling that forces
the components of $U$ and $B$ to decay at the same rate.
However, in our simulations, the microjets appear in all expanding runs, with or
without mean magnetic field, provided expansion is strong enough (i.e.
$\epsilon\gtrsim1$) and the Alfv\'en effect is not too strong, which in the case
of run D for instance is true only when $K_x=0$: this is enough to generate the microjets in this case.
In run E where the mean field is oblique, becoming close to $45^0$ to the radial at 1~AU, 
on may fear that the Alfv\'en effect at $K_x=0$ is still large, due to the non-vanishing contributions of non zero 
$K_y$ leading to a large Alfv\'en frequency $\omega = K_y B_y^0$. However, the most energetic scales now have $K_y$ smaller than unity, due to the transverse stretching of the plasma volume, which leads to finally an effective decoupling of the $U$ and $B$ components.

\subsection{Turbulent vs linear damping rate}
The spectral anisotropy which we have studied in the previous section suggests that the cascade process
of expanding turbulence is not simple. To advance in this direction, we propose
not only to measure the damping rate globally (has done in section
\ref{sec:emerg-struct-real}), but also to reveal its anisotropic nature in Fourier space.
A fundamental result of early 1D models of solar wind turbulence
(e.g. \citealt{Tu_al_1984}) is that of a clear partition of Fourier space in two parts: the one dominated by linear
expansion, to which the energy injection scales belong, and the one dominated
by non-linear couplings. 
However we have seen that expansion induces an anisotropic evolution of
spectra, so if we know where to place the 3D boundary of the two domains in
Fourier space, we will be in a better position to predict how the
injection scale varies and how the efficiency of turbulent heating is ruled
by expansion.

At a given scale, the competition between linear and non-linear
coupling can be understood by comparing the expansion time to the
non-linear time (see  definitions in
eqs.~\ref{eq:characteristictimes_tnl}-\ref{eq:characteristictimes_te}).
The expansion time increases linearly with distance.
At a given wavenumber, the nonlinear time should also increase as the
amplitude of the fluctuation decreases with distance (either simply due
to the expansion, or due to expansion and turbulent dissipation).
In principle, a good measure of the boundary is provided by the radial decay rate, since
as we know the linear analysis predicts well-defined decay rates for the different components with distance.

We now compare the energy decay rates for the three runs B, D, E in Fourier space to localize how nonlinear/linear decay rates varies with scale and direction.
In particular we compute the radial decay rate, $\alpha$, for the total
(magnetic $+$ kinetic) energy between
two times $t_0$ and $t_1$: 
\begin{equation}
  \label{eq:decay_alpha_def}
 \frac{E_{tot}(t_1)}{E_{tot}(t_0)} = \left( \frac{R(t_1)}{R(t_0)}
 \right)^{-\alpha}  
\end{equation}
In run E gyrotropy is no longer guaranteed, so we plot in Fig.~\ref{fig8} the
isocontours $\alpha=1,2,3$ in the ($K_x,~K_y$) plane at $K_z=0$. 
Recall that all three runs B, D, E have the same 
expansion  but different mean magnetic field. When it is present (run D and E),
we also plotted the two thin lines corresponding to directions parallel and
perpendicular to $B_0$ at that time (in run E,
$B_0$ forms an angle of $\approx45^o$ with the $x$ axis).

We can see that all three runs show different decay rates at
different scales and directions: as expected, isotropy is never achieved.

To summarize, the isocontour $\alpha=1$ corresponds to the WKB decay rate
and reflects approximately the global symmetry of the system: a
radial symmetry for runs B and D, and a complex one for run E. In the latter, one can identify two symmetry axes: (i) the direction perpendicular to the radial
(ii) the direction perpendicular to the mean field. The presence of these two
symmetry axes was already highlighted in the observational study of 
\citet{Saur_Bieber_1999}.
In all cases, the faster decay is always along the radial axis: gradients perpendicular to the radial dissipate less
easily than gradients along the radial.
Most interestingly, as already seen from fig.~\ref{fig0}, runs with nonzero mean field decay faster with distance. This may be explained because 
the mean field leads to a larger Alfv\'en
coupling and thus forces the otherwise linearly conserved quantities ($B_{y,z}$, $U_x$) to
follow the intermediate $1/R$ energy decay rate.

%We are currently developing a more accurate
%method involving the computation of the cascade rate thorugh the Politano-Pouquet law \citep{PP98} adapted to the EBM \citep{Hellinger_al_2013}. 
%we still have to understand how injection occurs in an expanding system. In fact, injection scales are governed by non-WKB
%and WKB expansion (the latter including a reduction of nonlinear transfer in
%field-parallel direction) in a non trivial way, also inducing polarization anisotropies that affect the development of the cascade. 
%
\begin{figure}[tbh]
\begin{center}
\includegraphics [width=0.9\linewidth]{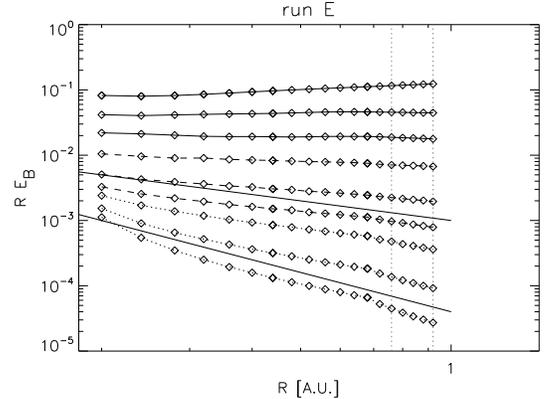}
\caption{
Run E. Radial decrease vs heliocentric distance $R$ of the 1D magnetic energy spectrum $E_B(K_x)$ at different radial wavenumbers:
$K_x=1,2,4,8,16,24,32,48,64$.
Curves are compensated by a $1/R$ decrease. 
Solid lines: $1/R^2$ and $1/R^3$ decay laws.
Vertical dotted lines mark the two distances between which the radial decay rate is computed and represented in the previous fig.~\ref{fig8}
%One can see the non-self-similar evolution of the spectrum above the inertial range ($k>8$), as observed in \citet{Bavassano_al_1982}. 
}
\label{fig7}
\end{center}
\end{figure}

\begin{figure*}[th]
\begin{center}
\includegraphics [width=0.4\linewidth]{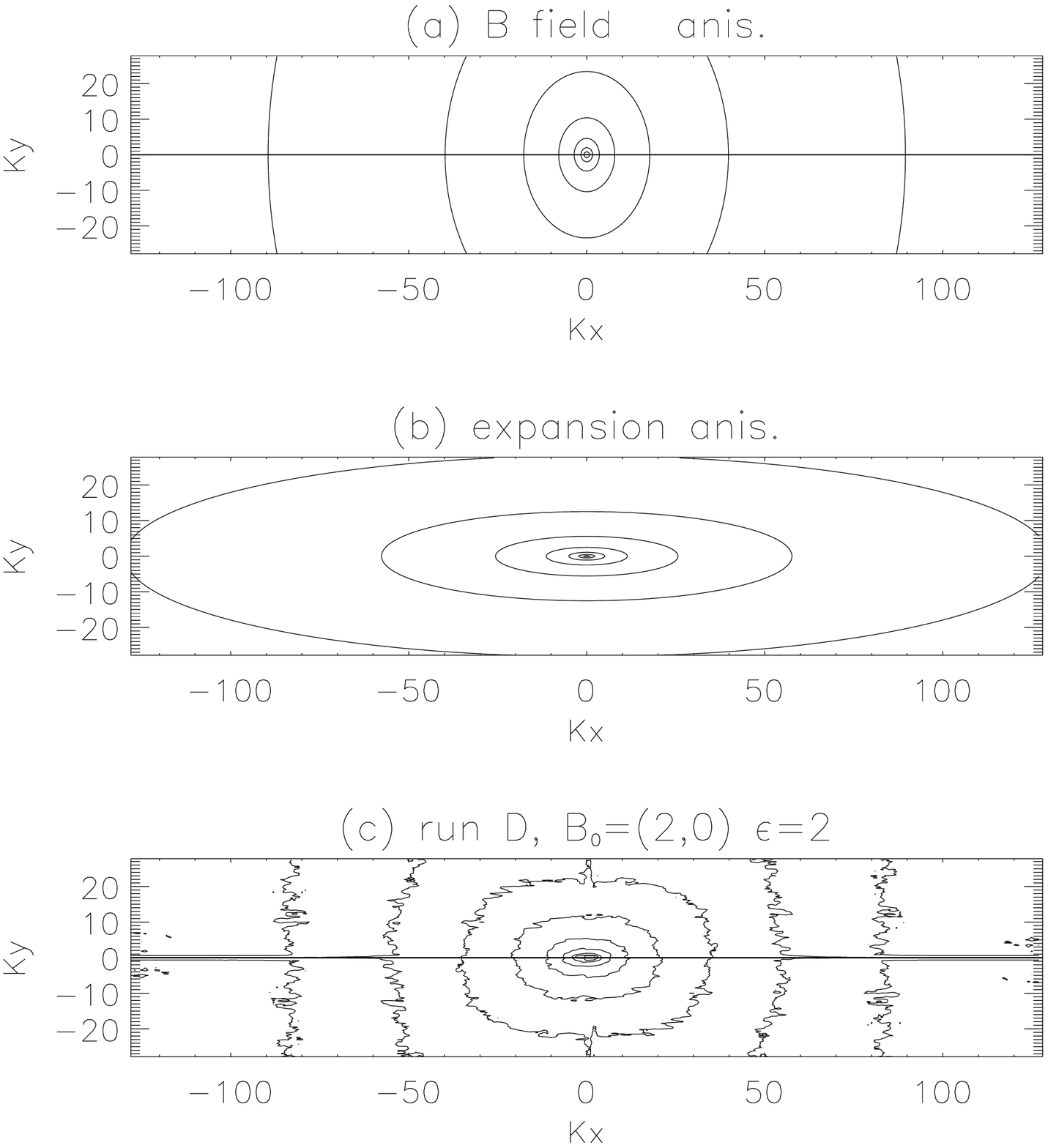}
\includegraphics [width=0.4\linewidth]{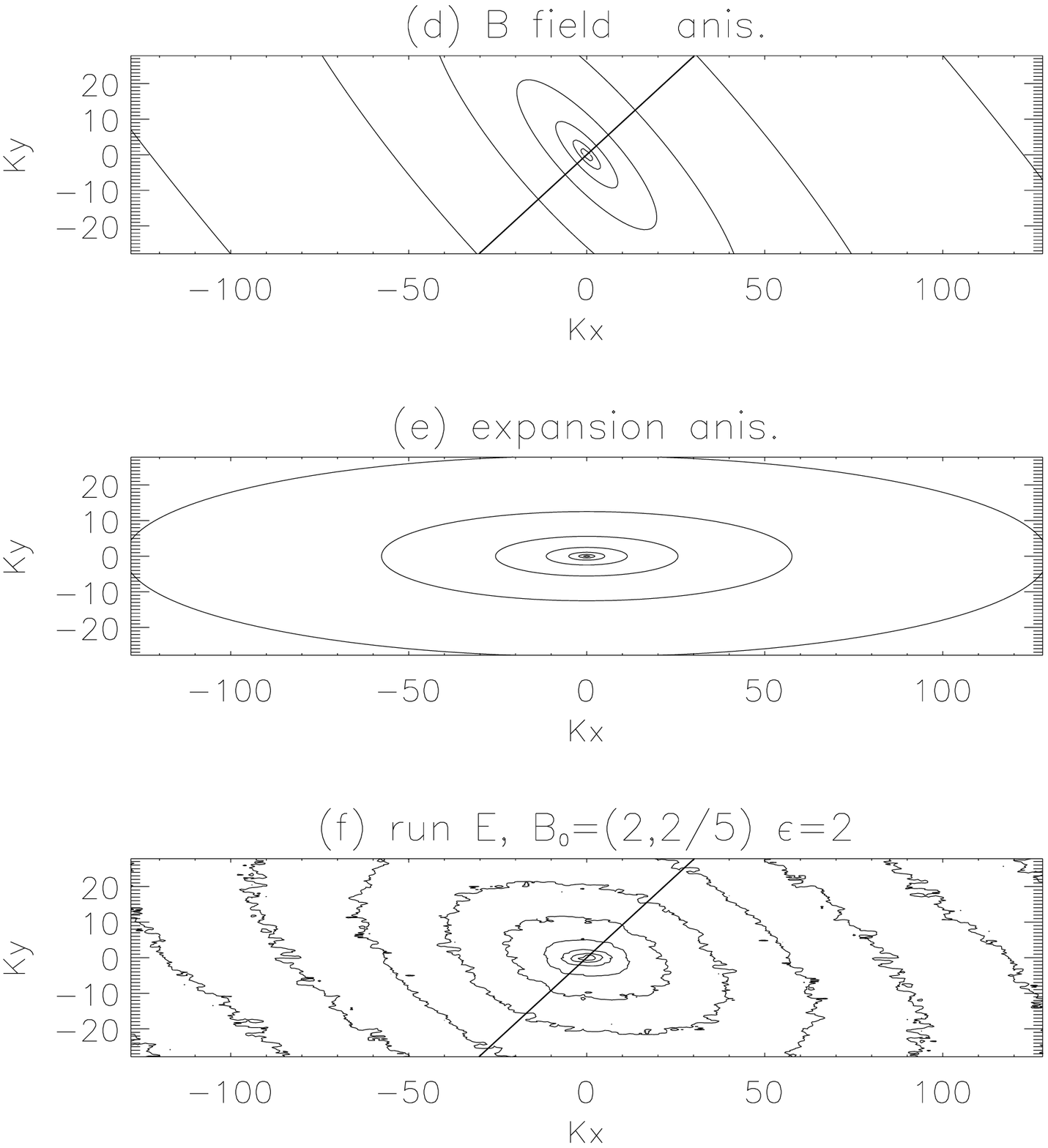}
\caption{Runs D and E. Mechanism for anisotropy formation.
Cut in the ($K_x,~K_y,~0$) plane of the 3D magnetic energy
spectrum.
Left column: case with radial mean field. Right column: case with oblique mean field.
Top panels: sketch of spectra perpendicular to mean
field (straight lines indicate the mean field direction) driven by the solar nonlinear terms.
Mid panels: sketch of the kinematic contraction of the spectrum due to the sole linear expansion.
Bottom panels: 
True spectral isocontours for run D (left) and E (right) at time $t=1.8~t_{NL}$. 
%Run E shows important deviations from gyrotropy due to the presence of two axes of symmetry. Strong deviations are found at small scales ($K>20$), where expansion and magnetic field anisotropies mix. At the largest scales, spectra are symmetric around the radial axis, because expansion dominates.
}
\label{fig6}
\end{center}
\end{figure*}
To compare with observational data,
in fig.~\ref{fig7}  
we plot the radial decay of magnetic energy density at
different radial wavenumbers (increasing from top to bottom) for run E.
The curves are compensated by $1/R$ which is the expected WKB decay
for large scales subject to Aflv\'enic coupling.
We can distinguish to wavenumber intervals. 
Large parallel scales ($2 \le K_x \le 8$) follow the $1/R$ WKB law, while smaller scales decay as $E\propto 1/R^2$ in the short interval $24 \le K_x \le 32$.
This behavior is akin to that reported by \citet{Bavassano_al_1982} who analyzed Helios data of
fast streams emanating from an equatorial extension of a coronal hole. 
In the data, the WKB scaling is found for a frequency band belonging to the
$1/f$ range of the spectrum, while higher frequency bands in the $1/f^{5/3}$
range have a $1/R^2$ decay.

In fig.~\ref{fig7} the wavenumber interval with $1/R$ and $1/R^2$ scaling laws is short. 
This reflects the fact that the spectrum (not shown) displays a smooth transition from the $1/f$ branch to the $1/f^{5/3}$ branch, at variance with the sharper spectral break found in observations. 
The origin of this difference probably lies in the limited range of scales
available in our simulations, which makes difficult to obtain two clearcut
power law ranges in a such a small interval of scales.

\subsection{Spectral anisotropy}

After exploring the anisotropy of the decay rates in Fourier space, let us return to the spectral anisotropy, and see if we can generalize the analysis given for the radial mean field to the oblique field case.
We thus compare runs D and E.

Using as a diagnostic tool the 1D reduced spectra (not shown), we find that run E has basically the same
component anisotropy as run D (fig.~\ref{fig2}d). Nevertheless, one expects
differences in spectral anisotropy due to the fact that the two axis of symmetry are no longer parallel in run E.
To see these differences in detail, we consider cuts of the 3D energy
spectrum in the $K_x,~K_y$ plane at $K_z=0$, since gyrotropy is not expected for run E.
In fig.~\ref{fig6} we combine such cuts with sketches that show how the ideas developed for run D can be extended to cope with the oblique case of run E.

As we know, energy tends to cascade in directions
perpendicular to the mean field in Fourier space, resulting in an
elongation of contours in the direction perpendicular to the mean field. Sketches of such spectra 
are plotted in the top panels (a) and (d) of fig.~\ref{fig6}, where the thick line indicates the 
direction of the mean magnetic field.

We also know that the linear kinematic expansion leads to a contraction in Fourier space on the $K_x$ axis, thus showing another, well-defined
anisotropy with symmetry axis along the radial direction.
A sketch of such spectra induced by expansion of an initial isotropic spectrum is
plotted in the middle panels (b) and (e) of fig.~\ref{fig6}.

On the bottom panels of the figure we finally show the true spectra, at time 1.8 (corresponding to a
heliocentric distance of about 1~AU). 
They result from the combinination of the two previous transformations.
In run D (panel c), the magnetic field is aligned with the radial, so there is just one symmetry axis. 
The two
anisotropies work against each other, with the cascade pushing the isocontours
in the $K_y$ direction and expansion pulling toward smaller $K_y$. The
final  spectrum is almost isotropic, although a careful inspection actually shows that large scales ($k<10$) follow the anisotropy induced by expansion, while small
scales tend to recover the anisotropy induced by the magnetic field. 

In the case with oblique mean field (panel
f), we can distinguish the two different axes of symmetry given by the radial direction (expansion) and by the direction of the magnetic field
(the line at approximately $45^o$). One clearly sees that at large scales the
symmetry axis is the radial, as should be the case for the expansion-dominated
system; at smaller scales the isocontours elongate progressively along the
oblique axis perpendicular to the mean field, while still keeping some of the
raidal anisotropy. Thus, expansion
affects also the anisotropy of ``turbulent'' scales, and not only of the
large energy-containing scales. 
We recall that a clear deviation from gyrotropy has been found in observations in the early
work by \citet{Saur_Bieber_1999} and more recently in the work by
\citet{Narita_al_2010}. 

As a final remark on spectral anisotropy, we mention that 
the tendency of an increasing radial anisotropy with increasing heliocentric
distances found in our simulation is in apparent contraddiction with observations. 
In the solar wind, the measure of correlations scales in field-parallel and
field-perpendicular directions shows a tendency of fluctuations to align in the
the plane perpendicular to the mean field (e.g. \citealt{Dasso_al_2005, Ruiz_al_2011}. We rather observe the opposite
tendency in run D, with radial magnetic field. However, in run E with oblique
mean field, the initial isotropic spectrum evolves into a spectrum with more
energy in the field-perpendicular direction (fig.~\ref{fig6}f), consistent with
the above mentioned observations. Thus, having a rotating mean magnetic field
(i.e. two distinct axis of
symmetry) appears to be fundamental in determining the evolution of
spectral anisotropy with distances, since rotations allows turbulent scales to
partially escape the effect of expansion.

\subsection{Component anisotropy: mechanism for the magnetic and perpendicular excess \label{structure}}
We want to explain here why expansion enhances the magnetic excess when it exists (the zero mean field case) and why it generates in all cases (that is, whatever the mean field) an excess of the perpendicular components.

Consider the zero mean field case first. As already said, we assume that the excess of magnetic energy in non expanding runs with zero mean field is due to the local dynamo effect which transfers energy from kinetic to magnetic components (see \cite{Muller:2005jp}),
with the Alfv\'en effect allowing to reach a balance with the dynamo, so leading to a definite value of the magnetic excess at each scale, the Alfv\'en effect dominating completely at the dissipative scales.
\begin{figure}[h]
\begin{center}
\includegraphics [width=\linewidth]{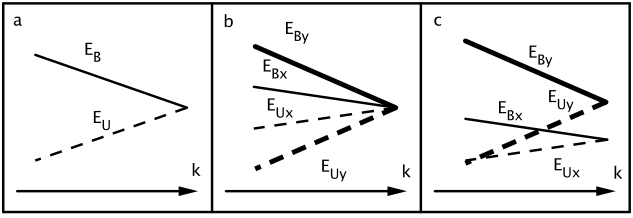}
\caption{Schematics of component anisotropy mechanisms. See text.
}
\label{fig:pola_aniso_mecha}
\end{center}
\end{figure}
This is summarized in fig.\ref{fig:pola_aniso_mecha}a which shows schematically the respective magnetic and velocity spectra, summarizing the regime of run A.

In the expanding case (run B), one observes a remarkable but puzzling ordering of the spectra
for all scales visible in fig.~\ref{fig2}b:
\be
U_x < B_x < U_y < B_y
\label{ineqspe}
\ee
Expansion here introduces a selective decay of the different
  components. We can write schematically the equations combining the Alfv\'en-dynamo (AD) effect
 and the damping terms of eqs.~\eqref{equ} and \eqref{eqB} for the residual
  energy $E_B-E_U$ of $x$ or $y,z$ components:
  \begin{eqnarray}
    \label{eq:ER_x_yz}
    \frac{d(E_{Bx}-E_{Ux})}{dt} &=& AD -\frac{2E_{Bx}}{t_{exp}} \\
    \frac{d(E_{By}-E_{Uy})}{dt} &=& AD +\frac{2E_{Uy}}{t_{exp}}
  \end{eqnarray}
with the expansion time $t_{exp}$ defined in eq.~\eqref{eq:characteristictimes_te}. Because components are damped differently (damped for $B_x, U_y,
U_z$, not for $B_y, B_z, U_x$), depending on whether this damping
fights against or in favor of the magnetic excess, we will
thus expect a larger magnetic excess for the perpendicular components than for the parallel one.
This is schematized in fig. \ref{fig:pola_aniso_mecha}b. 

Finally, to understand why the total (kinetic + magnetic) energy is larger for a perpendicular than a parallel component, one must consider total energy.
Local dynamo and Alfv\'en terms only appear as transfers
  between magnetic and kinetic fields, and are therefore absent from the total energy
  equation which can be written as:
  \begin{eqnarray}
    \label{eq:E_tot_x_yz}
    \frac{d(E_{Bx}+E_{Ux})}{dt} &=& -\frac{2E_{Bx}}{t_{exp}} 
\label{select1}
\\
    \frac{d(E_{By}+E_{Uy})}{dt} &=& -\frac{2E_{Uy}}{t_{exp}}
\label{select2}
 \end{eqnarray}
The damping rate for the total energy of the $x$ component is based on
$E_{Bx} \approx E_{tot,x}$ whereas the damping rate for the $y$
component of the total energy is based on $E_{Uy} \ll
E_{tot,y}$. This creates a gap between $x$ and $y,z$ components:
$E_{tot,y} > E_{tot,x}$, as illustrated on
fig. \ref{fig:pola_aniso_mecha}c. We have thus explained the ordering of component spectra of run B.

The case of run D is simpler: the Alfv\'en effect (equipartition between B and U) dominates completely the dynamo effect, so equipartition holds between U and B fields, and only the selective decay due to expansion is at work (eqs.~\ref{select1}-~\ref{select2}), so that the perpendicular components dominate.

\begin{figure}[t]
\begin{center}
% CORRECT FOR TWO-COLUMNS
%\includegraphics [width=\linewidth]{fig-composite2.eps}\\
%\includegraphics [width=0.49\linewidth]{figB-poly1.eps}
%\includegraphics [width=0.49\linewidth]{figB-poly2.eps}
% ADAPTED FOR ONE-COLUMN
\includegraphics [height=0.45\linewidth]{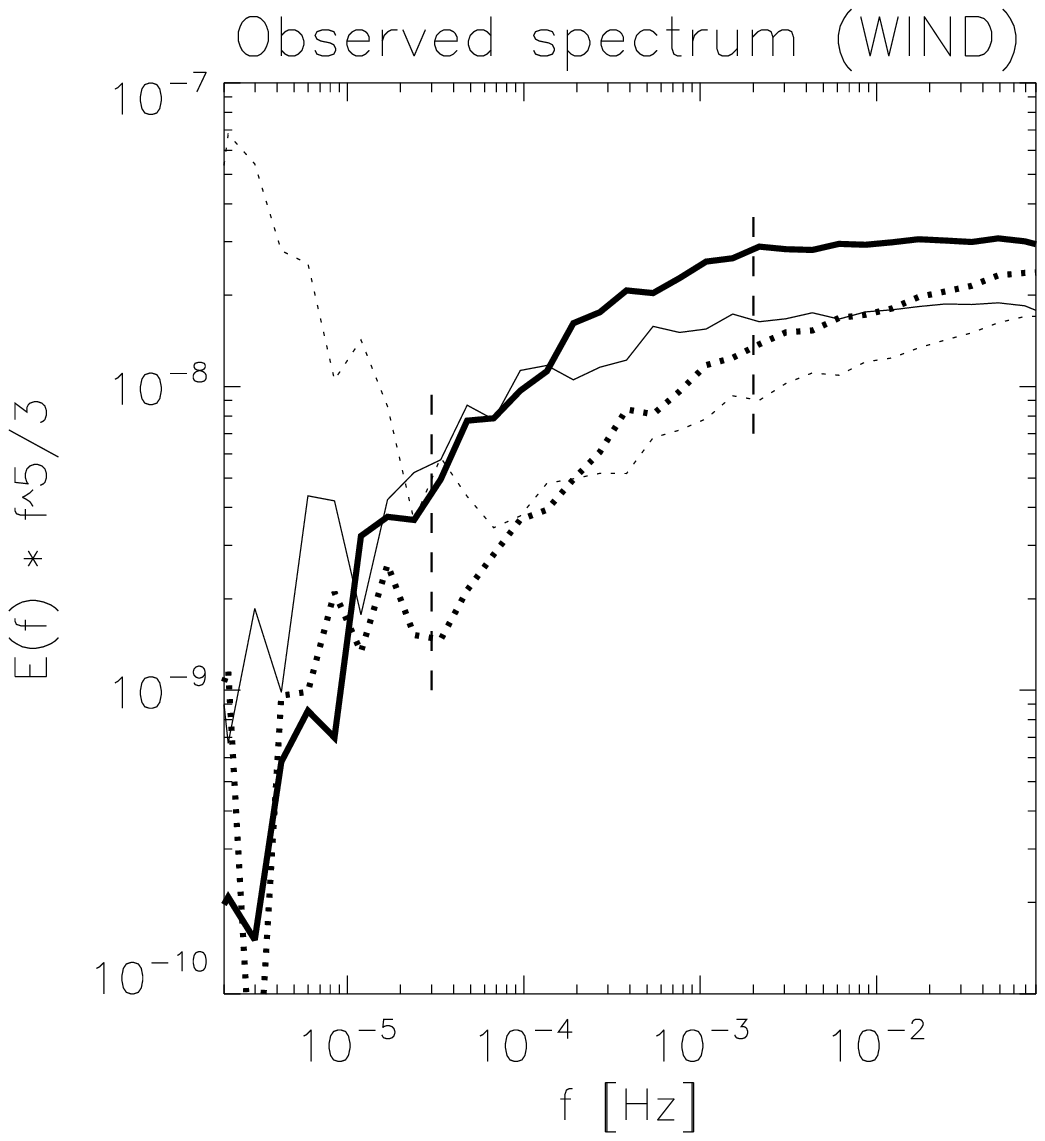}\\
\includegraphics [width=0.41\linewidth]{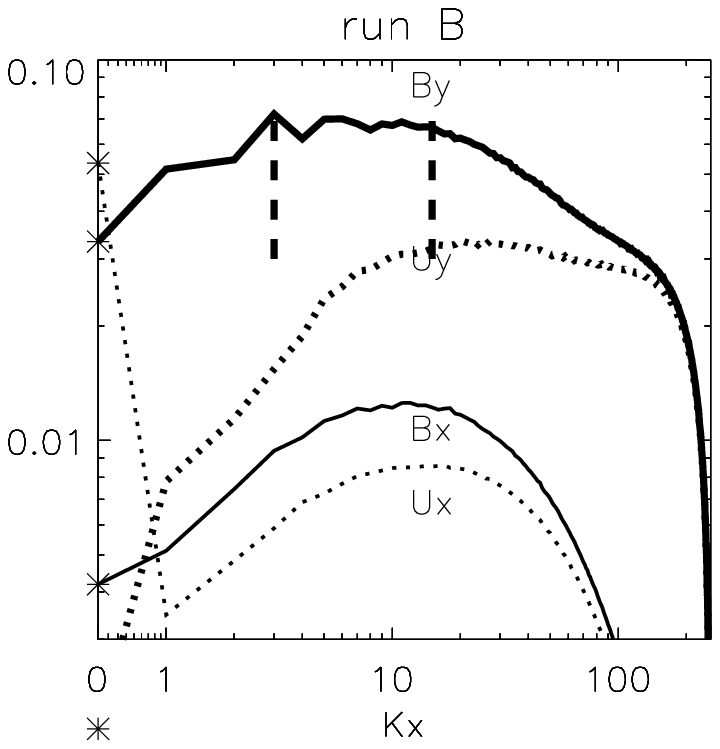}
\includegraphics [width=0.41\linewidth]{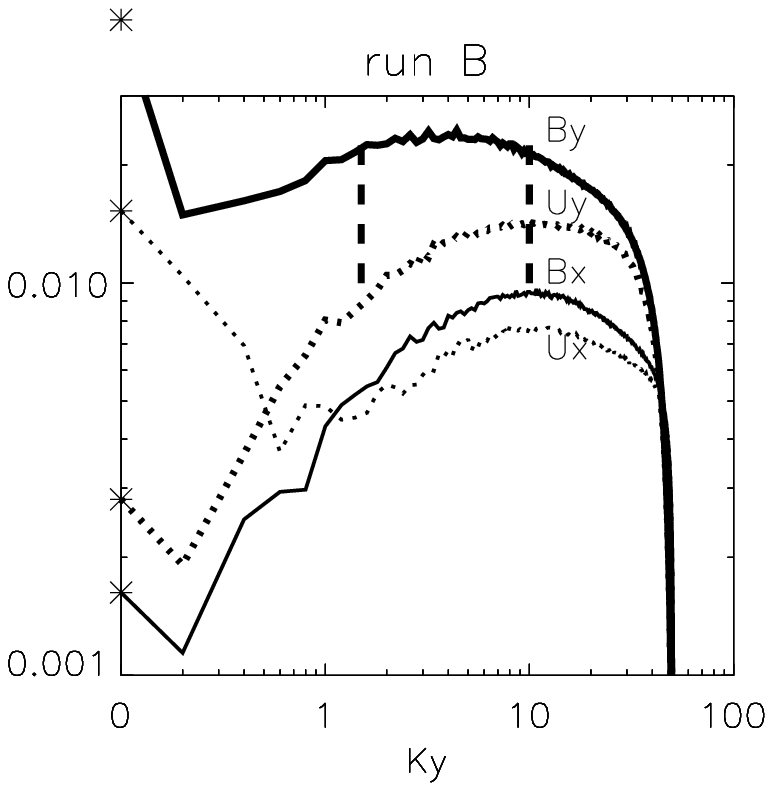}
%\caption{Polarization spectra: comparing frequency spectra $E(f)$ in Wind mission and simulation results of run B: reduced spectra along the radial ($K_x$) and perpendicular ($K_y$) directions, \textit{including the energy density at $K_x=0$ (or $K_y=0$)}.
\caption{Spectra, component by component. Comparison between the observed
frequency spectra (top) and the wavenumber spectra of run B 
(bottom).
Top panel: frequency spectra compensated by $f^{-5/3}$ obtained from Wind
mission during April 1995-July 1995.
Solid lines: magnetic field. Dashed lines: velocity field. Thin lines: radial component (R). Thick lines: component perpendicular to the ecliptic (N).
Bottom panels: spectra of run B at time $t=2$, same line styles as in fig.~\ref{fig2}.
Left: radial spectra $E(K_x)$ compensated by $K_x^{-5/3}$. Right:
perpendicular (to the radial) spectra $E(K_y)$ compensated by $K_y^{-5/3}$.
In the bottom figures, the vertical dashed bars mark the extent of the
inertial range, with scaling $K^{-5/3}$. In the top figure instead, the
vertical dashed bars mark the extent of the $f^{-1}$ fossil range. On the right
most and on the left most sides one finds respectively the inertial range and the very large
scales, the latter including longitudinal structures as the stream structure.}
\label{fig2_data}
\end{center}
\end{figure}

How do the computed spectra compare with observations?
We consider the spectra for each component in solar wind data at 1~AU. 
In fig.~\ref{fig2_data}a we plot the spectra
measured by the WIND mission, with spectra compensated by $f^{5/3}$.
The period considered is around the minimum of solar activity, from April to
July 1995, containing a mixed population of slow and fast streams.
Components radial and normal to the ecliptic are indicated by respectively thin and thick lines, velocity and magnetic energy (in energy per unit mass) by solid and dotted lines.
One can distinguish three spectral ranges in this figure: (i) the $f^{-5/3}$ range for the magnetic field on the right of the right vertical dashed bar ($f > 2~10^{-3}$~Hz), (ii) the $f^{-1}$ central range between the two dashed bars, (iii) a large frequency range ($f < 3~10^{-5}$~Hz or periods larger than 9 hours) where one finds the dominant energy specific of the largest scales, namely the radial stream structure ($U_r$) that characterizes the large-scale solar wind, showing the imprint of the latitudinal and longitudinal magnetic topology of the solar corona. 

Starting from the higher frequencies, one sees an excess of the non-radial component, and an excess of the magnetic energy that grows as frequency decreases. The magnetic excess continues to increase when entering the $k^{-1}$ range, but saturates in the middle of this range, and the dominant energy becomes then that of the radial velocity, $U_r$.
In order to allow for a detailed comparison, we have added (bottom left panel, fig.~\ref{fig2_data}b) the $E(K_x)$ energy spectra of run B, including the energy in the perpendicular direction at $K_x=0$. Fig.~\ref{fig2_data}c is the same but for the $E(K_y)$ spectra.
We see that the range between $K\simeq 1$ and $K>15$ compares favorably with the range $3~10^{-5}\le f \le  8~10^{-2}$ that encompasses the $f^{-1}$ and inertial ranges in observational data.
The main points of agreements are (i) the systematic magnetic excess, (ii) the systematic excess in perpendicular components, (iii) the growth of energy in the radial velocity component at the largest scales.
In the simulations of run B, the radial streams are completely devoid of radial structures, so they appear only in the $K_x=0$ modes, while they appear also at the large non zero $K_y>0$ modes in the transverse spectra $E(K_y)$. In the observational data, it is worth to mention that the low frequency range in which the radial stream structure begins to appear actually mixes radial, longitudinal and latitudinal structures. 
This mixing is absent in run B that lacks of the initial coronal rotation.

There are however points of disagreement (i) the relative excess of the perpendicular components is much larger in the simulations than in the real wind (ii) the exchange of ordering between the $B_y$ and $U_x$ components that is observed in the data at large scales is not found in the simulations, where the largest scales are dominated by both $B_y$ and $U_x$.
A plausible explanation for the latter point is that the true initial conditions in the solar wind at $0.2~AU$ very probably show already some dominance of the $U_x$ component, while we considered instead equipartition between all degrees of freedom in our simulations.
The first point is a consequence of the second.
Assuming as usual a direct cascade, any increase of the large scale level of the $U_x$ component will
automatically lead, due to the Alfv\'en coupling, to a larger level of both the $U_x$ and $B_x$ components at smaller scales.

Two last remarks are in order. First, the magnetic excess shown by our observational data is a clear signature of the slow streams, not of the fast streams: the latter show a reduced or no magnetic excess \citep{Grappin:1991tr}. This is the reason why we compared the observational spectra with that obtained in run B.
Second, the mechanism that we propose here to explain the specific features of the component spectra found in the wind and in our data extends to the expanding wind the mechanism proposed by \cite{1983A&A...126...51G} and \cite{Muller:2005jp}.
The mechanism involves a balance at all scales between the local dynamo effect and the Alfv\'en effect.
In order to lead to the observed ordering and scaling (i.e., a large magnetic excess at large scale, and equipartition at small scales), the Alfv\'en effect must be dominant on the dynamo effect at all scales.
%This strongly suggests weak nonlinear coupling, not strong nonlinear coupling, although the magnetic (and total) energy scaling is $k^{-5/3}$. This issue is the object of work in progress.

\section{Conclusion}

We have studied the evolution of turbulence using for the first time full 3D
MHD simulations that are able to account for  
all basic physical effects due to the anisotropic expansion of a
plasma volume embedded in a spherically expanding wind at
constant speed (the Expanding Box Model, EBM). 
The main results can be summarized in the following three points.

(1) The \textit{spectral anisotropy} is mainly
due to the side-way stretching of the wind, that controls the dynamic at
injection scales and acts at all scales, thus influencing \textit{twice} the
small scale anisotropy. At a distance of approximately 1~AU, the final anisotropy is determined by the two axis of symmetry of the
problem, the radial axis (expansion) and the mean magnetic field axis
(turbulence), with more energy residing in radial wave-vectors. In the
case of radial mean field, the spectral anisotropy at a given
heliocentric distance can be qualitatively predicted by knowing the initial
mean field and rms amplitude of fluctuations (see eq.~\ref{eq:defA}).

(2) We identified three main mechanisms that are responsible for the development of \textit{component anisotropies} in our simulations.
The linear damping due to expansion 
rules the decay of 2D modes, that is the largest parallel scales
with $K_x=0$ (2D is with respect to the radial direction). At all other scales, the component anisotropy results from a competition between expansion, non-linear stretching of the magnetic field (local dynamo), and the Alfv\'en
relaxation effect (either based on the global or local mean fields). 
This combination is able to explain the component anisotropy in
simulations, and several features of the component anisotropy observed in the solar wind.
However a discrepancy remains: the excess of transverse magnetic field that is not found in solar wind data. 

(3) The selective decay of components combines with the spectral anisotropy
to yield non-turbulent \textit{radial streams} that resemble the observed
microjets \citep{McComas_al_1995,Neugebauer_al_1995}. 
We anticipate that they differ in some aspect from the observations, and
defer to future work a careful analysis of these structures.

We considered initial spectra that are isotropic and at equipartition, between
kinetic and magnetic energy and among components. 
These initial conditions are not entirely representative of turbulence in the fast or slow wind, however we are able to recover most of the observed features, in particular the component anisotropy. 
This proves that the turbulence properties observed at 1~AU are at least
partially due to the evolution of turbulence during its transport in the
heliosphere, and are not a simple remnant of the initial coronal turbulence
close to the Sun.

It is worth noting that some of our results can be used to constrain the
unknown turbulence spectrum in the inner heliosphere. 
Our microjets are a remnant of initial
conditions, they emerge because of the selective decay due to expansion. This suggests that they probably
have a solar origin and offers an explanation (expansion) for their survival.
On the other hand, the difference between simulated and observed spectral
anisotropy suggests a weaker initial transverse magnetic field in the solar
wind compared to our isotropic initial condition.

Further work is however needed. 
%to fully understand the influence of coronal
%spectrum on the following evolution in the heliosphere. 
Based on the measure of radial decay of energy we were able to roughly identify the energy containing scales (fig.~\ref{fig8}) that appear to be anisotropic, with the anisotropy being determined by both expansion and nonlinear interactions.
We are currently working on a better identification of the energy containing
scales by direct computation of the cascade rate through the Politano-Pouquet
law \citep{PP98} in a version adapted to the EBM  
\citep{Hellinger_al_2013}. This will allow us to understand the
dynamic of injection scales and hence the imprint of coronal turbulence
on the heliospheric spectra. 
Concerning dissipation, we also plan to modify the dissipative terms so as to allow a correct  transfer of the dissipated turbulent energy into the internal energy of the plasma, 
while at the same time preventing a too sharp drop of the Reynolds number with distance.

We further plan to implement more realistic initial conditions, such as
component and spectral anisotropy, or imbalance of outward-inward Elsasser fields. 
%The latter in particular could be fundamental in reproducing the observed spectral break in fast streams, since it is another source of quenching for the cascade and could efficiently maintain the low frequency $1/f$ spectrum. 
Finally, we also plan to analyze our
data through simulated flybys and perform a local analysis to compare directly
with more recent observations. 

%\begin{acknowledgments}
%\textit{Acknowledgments}
\acknowledgments
We thank warmly Marco Velli for several fruitful discussions and Fabrice Roy who
has parallelized the 3D EBM code.
The research leading to these results has received partial funding from the
European Commission's Seventh Framework Programme (FP7/2007-2013) under the
grant agreement SHOCK (project number 284515) and from the Interuniversity
Attraction Poles Programme initiated by the Belgian Science Policy Office (IAP
P7/08 CHARM).
This work was performed using HPC resources from GENCI-IDRIS (Grant 2014-040219)
and CINECA (ISCRA class C project HP10CDO94O).
%\end{acknowledgments}

%\bibliography{poster2}

\begin{thebibliography}{45}
\expandafter\ifx\csname natexlab\endcsname\relax\def\natexlab#1{#1}\fi

\bibitem[{Bavassano {et~al.}(1982)Bavassano, Dobrowolny, Mariani, \&
  Ness}]{Bavassano_al_1982}
Bavassano, B., Dobrowolny, M., Mariani, F., \& Ness, N.~F. 1982, J. Geophys.
  Res., 87, 3616

\bibitem[{Belcher \& Davis(1971)}]{Belcher:1971cn}
Belcher, J.~W. \& Davis, L. 1971, Journal of Geophysical Research, 76, 3534

\bibitem[{Bruno {et~al.}(2007)Bruno, D'Amicis, Bavassano, Carbone, \&
  Sorriso-Valvo}]{Bruno:2007tw}
Bruno, R., D'Amicis, R., Bavassano, B., Carbone, V., \& Sorriso-Valvo, L. 2007,
  Annales Geophysicae, 25, 1913

\bibitem[{Chandran {et~al.}(2011)Chandran, Dennis, Quataert, \&
  Bale}]{Chandran:2011p2723}
Chandran, B. D.~G., Dennis, T.~J., Quataert, E., \& Bale, S.~D. 2011, The
  Astrophysical Journal, 743, 197

\bibitem[{Coleman(1968)}]{1968ApJ...153..371C}
Coleman, P. J.~J. 1968, Astrophysical Journal, 153, 371

\bibitem[{Cranmer \& van Ballegooijen(2005)}]{CB05}
Cranmer, S.~R. \& van Ballegooijen, A.~A. 2005, The Astrophysical Journal
  Supplement Series, 156, 265

\bibitem[{Cranmer {et~al.}(2007)Cranmer, van Ballegooijen, \&
  Edgar}]{2007ApJS..171..520C}
Cranmer, S.~R., van Ballegooijen, A.~A., \& Edgar, R.~J. 2007, The
  Astrophysical Journal Supplement Series, 171, 520

\bibitem[{Dasso {et~al.}(2005)Dasso, Milano, Matthaeus, \&
  Smith}]{Dasso_al_2005}
Dasso, S., Milano, L.~J., Matthaeus, W.~H., \& Smith, C.~W. 2005, The
  Astrophysical Journal, 635, L181

\bibitem[{Frisch(1995)}]{1995tlan.book.....F}
Frisch, U. 1995, Turbulence. The legacy of A. N. Kolmogorov.

\bibitem[{{Grappin}(1996)}]{Grappin_1996}
{Grappin}, R. 1996, in American Institute of Physics Conference Series, Vol.
  382, American Institute of Physics Conference Series, ed. D.~{Winterhalter},
  J.~T. {Gosling}, S.~R. {Habbal}, W.~S. {Kurth}, \& M.~{Neugebauer}, 306--309

\bibitem[{Grappin {et~al.}(1983)Grappin, L{\'e}orat, \&
  Pouquet}]{1983A&A...126...51G}
Grappin, R., L{\'e}orat, J., \& Pouquet, A. 1983, Astronomy and Astrophysics
  (ISSN 0004-6361), 126, 51

\bibitem[{Grappin \& Velli(1996)}]{Grappin_Velli_1996}
Grappin, R. \& Velli, M. 1996, J. Geophys. Res., 101, 425

\bibitem[{Grappin {et~al.}(1991)Grappin, Velli, \& Mangeney}]{Grappin:1991tr}
Grappin, R., Velli, M., \& Mangeney, A. 1991, Annales Geophysicae (ISSN
  0939-4176), 9, 416

\bibitem[{Grappin {et~al.}(1993)Grappin, Velli, \& Mangeney}]{GVM93}
---. 1993, Phys. Rev. Lett., 70, 2190

\bibitem[{Hellinger {et~al.}(2013)Hellinger, Tr{\'a}vn{\'\i}{\v c}ek, Matteini,
  \& Velli}]{Hellinger_al_2013}
Hellinger, P., Tr{\'a}vn{\'\i}{\v c}ek, P.~M., Matteini, L., \& Velli, M. 2013,
  Journal of Geophysical Research, 118, 1

\bibitem[{Lionello {et~al.}(2014)Lionello, Velli, Downs, Linker, Miki{\'c}, \&
  Verdini}]{Lionello:2014p3023}
Lionello, R., Velli, M., Downs, C., Linker, J.~A., Miki{\'c}, Z., \& Verdini,
  A. 2014, The Astrophysical Journal, 784, 120

\bibitem[{MacBride {et~al.}(2008)MacBride, Smith, \&
  Forman}]{2008ApJ...679.1644M}
MacBride, B.~T., Smith, C.~W., \& Forman, M.~A. 2008, The Astrophysical
  Journal, 679, 1644

\bibitem[{Matsumoto \& Suzuki(2012)}]{Matsumoto:2012p2665}
Matsumoto, T. \& Suzuki, T.~K. 2012, The Astrophysical Journal, 749, 8

\bibitem[{Matteini {et~al.}(2006)Matteini, Landi, Hellinger, \&
  Velli}]{Matteini:2006jn}
Matteini, L., Landi, S., Hellinger, P., \& Velli, M. 2006, Journal of
  Geophysical Research, 111, 10101

\bibitem[{Matthaeus {et~al.}(1990)Matthaeus, Goldstein, \&
  Roberts}]{Matthaeus_al_1990}
Matthaeus, W.~H., Goldstein, M.~L., \& Roberts, D.~A. 1990, Journal of
  Geophysical Research (ISSN 0148-0227), 95, 20673

\bibitem[{Matthaeus {et~al.}(1999)Matthaeus, Zank, Smith, \&
  Oughton}]{Matthaeus:1999p630}
Matthaeus, W.~H., Zank, G.~P., Smith, C.~W., \& Oughton, S. 1999, Phys. Rev.
  Lett., 82, 3444

\bibitem[{McComas {et~al.}(1995)McComas, Barraclough, Gosling, Hammond,
  Phillips, Neugebauer, Balogh, \& Forsyth}]{McComas_al_1995}
McComas, D.~J., Barraclough, B.~L., Gosling, J.~T., Hammond, C.~M., Phillips,
  J.~L., Neugebauer, M., Balogh, A., \& Forsyth, R.~J. 1995, Journal of
  Geophysical Research, 100, 19893

\bibitem[{Montgomery {et~al.}(1978)Montgomery, Turner, \&
  Vahala}]{1978PhFl...21..757M}
Montgomery, D., Turner, L., \& Vahala, G. 1978, Physics of Fluids, 21, 757

\bibitem[{M{\"u}ller \& Grappin(2005)}]{Muller:2005jp}
M{\"u}ller, W.-C. \& Grappin, R. 2005, Physical Review Letters, 95

\bibitem[{Narita {et~al.}(2010)Narita, Glassmeier, Sahraoui, \&
  Goldstein}]{Narita_al_2010}
Narita, Y., Glassmeier, K.-H., Sahraoui, F., \& Goldstein, M.~L. 2010, Physical
  Review Letters, 104, 171101

\bibitem[{Neugebauer {et~al.}(1995)Neugebauer, Goldstein, McComas, Suess, \&
  Balogh}]{Neugebauer_al_1995}
Neugebauer, M., Goldstein, B.~E., McComas, D.~J., Suess, S.~T., \& Balogh, A.
  1995, Journal of Geophysical Research, 100, 23389

\bibitem[{Perez \& Chandran(2013)}]{Perez:2013p3017}
Perez, J.~C. \& Chandran, B. D.~G. 2013, The Astrophysical Journal, 776, 124

\bibitem[{Politano \& Pouquet(1998)}]{PP98}
Politano, H. \& Pouquet, A. 1998, Geophys. Res. Lett., 25, 273

\bibitem[{Pouquet {et~al.}(2010)Pouquet, Lee, Brachet, Mininni, \&
  Rosenberg}]{2010GApFD.104..115P}
Pouquet, A., Lee, E., Brachet, M.~E., Mininni, P.~D., \& Rosenberg, D. 2010,
  Geophysical and Astrophysical Fluid Dynamics, 104, 115

\bibitem[{Rappazzo {et~al.}(2005)Rappazzo, Velli, Einaudi, \&
  Dahlburg}]{Rappazzo_al_2005}
Rappazzo, A.~F., Velli, M., Einaudi, G., \& Dahlburg, R.~B. 2005, The
  Astrophysical Journal, 633, 474

\bibitem[{{Ruiz} {et~al.}(2011){Ruiz}, {Dasso}, {Matthaeus}, {Marsch}, \&
  {Weygand}}]{Ruiz_al_2011}
{Ruiz}, M.~E., {Dasso}, S., {Matthaeus}, W.~H., {Marsch}, E., \& {Weygand},
  J.~M. 2011, Journal of Geophysical Research (Space Physics), 116, 10102

\bibitem[{Sahraoui {et~al.}(2010)Sahraoui, Goldstein, Belmont, Canu, \&
  Rezeau}]{2010PhRvL.105m1101S}
Sahraoui, F., Goldstein, M.~L., Belmont, G., Canu, P., \& Rezeau, L. 2010,
  Physical Review Letters, 105, 131101

\bibitem[{Saur \& Bieber(1999)}]{Saur_Bieber_1999}
Saur, J. \& Bieber, J.~W. 1999, Journal of Geophysical Research, 104, 9975

\bibitem[{Sorriso-Valvo {et~al.}(2007)Sorriso-Valvo, Marino, Carbone, Noullez,
  Lepreti, Veltri, Bruno, Bavassano, \& Pietropaolo}]{2007PhRvL..99k5001S}
Sorriso-Valvo, L., Marino, R., Carbone, V., Noullez, A., Lepreti, F., Veltri,
  P., Bruno, R., Bavassano, B., \& Pietropaolo, E. 2007, Physical Review
  Letters, 99, 115001

\bibitem[{Tenerani \& Velli(2013)}]{Tenerani:2013p3022}
Tenerani, A. \& Velli, M. 2013, Journal of Geophysical Research: Space Physics,
  118, 7507

\bibitem[{{Tu}(1987)}]{1987SoPh..109..149T}
{Tu}, C.-Y. 1987, \solphys, 109, 149

\bibitem[{{Tu} \& {Marsch}(1993)}]{1993JGR....98.1257T}
{Tu}, C.-Y. \& {Marsch}, E. 1993, \jgr, 98, 1257

\bibitem[{Tu {et~al.}(1984)Tu, Pu, \& Wei}]{Tu_al_1984}
Tu, C.-Y., Pu, Z.-Y., \& Wei, F.-S. 1984, Journal of Geophysical Research (ISSN
  0148-0227), 89, 9695

\bibitem[{Vasquez {et~al.}(2007)Vasquez, Smith, Hamilton, MacBride, \&
  Leamon}]{2007JGRA..11207101V}
Vasquez, B.~J., Smith, C.~W., Hamilton, K., MacBride, B.~T., \& Leamon, R.~J.
  2007, Journal of Geophysical Research, 112, A07101

\bibitem[{Velli(1993)}]{Velli:1993to}
Velli, M. 1993, Astronomy and Astrophysics (ISSN 0004-6361), 270, 304

\bibitem[{{Velli} {et~al.}(1990){Velli}, {Grappin}, \&
  {Mangeney}}]{1990CoPhC..59..153V}
{Velli}, M., {Grappin}, R., \& {Mangeney}, A. 1990, Computer Physics
  Communications, 59, 153

\bibitem[{Verdini {et~al.}(2012)Verdini, Grappin, Pinto, \&
  Velli}]{2012ApJ...750L..33V}
Verdini, A., Grappin, R., Pinto, R., \& Velli, M. 2012, The Astrophysical
  Journal Letters, 750, L33

\bibitem[{Verdini \& Velli(2007)}]{Verdini_Velli_2007}
Verdini, A. \& Velli, M. 2007, The Astrophysical Journal, 662, 669

\bibitem[{Verdini {et~al.}(2009)Verdini, Velli, \& Buchlin}]{Verdini:2009ih}
Verdini, A., Velli, M., \& Buchlin, E. 2009, The Astrophysical Journal Letters,
  700, L39

\bibitem[{Verdini {et~al.}(2010)Verdini, Velli, Matthaeus, Oughton, \&
  Dmitruk}]{Verdini:2010et}
Verdini, A., Velli, M., Matthaeus, W.~H., Oughton, S., \& Dmitruk, P. 2010, The
  Astrophysical Journal Letters, 708, L116

\end{thebibliography}

\end{document}